\documentclass[pra, reprint, twocolumn, superscriptaddress, amsfonts, amssymb, amsmath]{revtex4-2}
\usepackage[english]{babel}

\usepackage{graphicx}

\graphicspath{{./Images/},{./ImagesAppendix/},
{./images/},{./imagesAppendix/}}

\usepackage{physics}
\usepackage{bm}
\usepackage{braket}
\usepackage{pgfplots}
\usepackage{array}
\usepackage{makecell}
\usepackage{amsmath}
\usepackage{tikz}
\usetikzlibrary{positioning, arrows.meta}
\usepackage{subfigure}
\usepackage{commath}
\usepackage{graphicx,bm}
\usepackage{verbatim}
\usepackage{flafter}
\usepackage{float}
\usepackage{varioref}
\usepackage{bbold}

\usepackage{mathtools}

\DeclarePairedDelimiter\floor{\lfloor}{\rfloor}

\usepackage{color}
\usepackage[colorlinks,citecolor=darkBlue,linkcolor=darkBlue,
urlcolor=blue,hyperindex]{hyperref}
\definecolor{darkBlue}{rgb}{0.08, 0.13, 0.4}

\definecolor{THc}{rgb}{0.9,0.3,0.2}

\begin{document}

%%%%%%%%%%%%%%%%%%%%%%%%%%%%%%%%%%%%%%%%%%%%%%
%%%  HEADING
%%%%%%%%%%%%%%%%%%%%%%%%%%%%%%%%%%%%%%%%%%%%%%

\title{Quantum superpositions of  current states in Rydberg-atom networks}
\author{Francesco Perciavalle}
\affiliation{Quantum Research Center, Technology Innovation Institute, P.O. Box 9639 Abu Dhabi, UAE}
\affiliation{Dipartimento di Fisica dell’Universit\`a di Pisa and INFN, Largo Pontecorvo 3, I-56127 Pisa, Italy}

\author{Davide Rossini}
\affiliation{Dipartimento di Fisica dell’Universit\`a di Pisa and INFN, Largo Pontecorvo 3, I-56127 Pisa, Italy}

\author{Juan Polo}
\affiliation{Quantum Research Center, Technology Innovation Institute, P.O. Box 9639 Abu Dhabi, UAE}

\author{Oliver Morsch}
\affiliation{CNR-INO and Dipartimento di Fisica dell’Universit\`a di Pisa, Largo Pontecorvo 3, 56127 Pisa, Italy}

\author{Luigi Amico}
\affiliation{Quantum Research Center, Technology Innovation Institute, P.O. Box 9639 Abu Dhabi, UAE}
\address{Dipartimento di Fisica e Astronomia ``Ettore Majorana" University of Catania, Via S. Sofia 64, 95123 Catania, Italy}
\affiliation{INFN-Sezione di Catania, Via S. Sofia 64, 95123 Catania, Italy}

\date{\today}

\begin{abstract}
Quantum simulation of many-body quantum systems using Rydberg-atom platforms has become of extreme interest in the last years. The possibility to realize spin Hamiltonians and the accurate control at the single atom level paved the way for the study of quantum phases of matter and dynamics. Here, we propose a quantum optimal control protocol to engineer current states: quantum states characterized by Rydberg excitations propagating in a given spatially closed tweezer networks. Indeed, current states with different winding numbers can be generated on demand. Besides those ones with single winding number,  superposition of quantum current states  characterized by more winding numbers can be obtained. The single current states are eigenstates of the current operator that therefore can define an observable that remains persistent at any time. In particular, the features of the excitations dynamics reflects the nature of current states, a fact that in principle can be used to characterize  the nature of the flow experimentally without the need of accessing high order correlators.
\end{abstract}

\maketitle

\section{Introduction}

Quantum simulation~\cite{feynman1982simulating,georgescu2014quantum, altman2021quantum} is one of the most promising approaches to understand quantum properties of matter using controllable systems. Useful platforms for quantum simulation include ultra-cold atoms~\cite{Greiner2002quantum, Bloch2012quantum, gross2017quantum}, trapped ions~\cite{Blatt2012quantum, Blatt2008entangled,monroe2021programmable}, superconducting circuits~\cite{schmidt2013circuit,lamata2018digital,fazio2020quantum} or Rydberg atoms trapped in optical tweezers~\cite{Weimer2010a,bernien2017probing,Browaeys2020many,Morgado2021quantum,monroe2021programmable, pause2024supercharged}.
Atomtronics, the emerging quantum technology  of guided flows of neutral matter~\cite{amico2005quantum,polo2023perspective,amico2022colloquium, Amico2021roadmap,pepino2021advances}, can provide a reference for new schemes in the field.   
In particular, the atomtronics logic can be generalized  to Rydberg atoms: instead of the matter-wave motion employed in conventional atomtronics, a controlled flow in terms of Rydberg excitations moving in circuits  comprised of suitable networks of tweezers~\cite{Endres2016atom,schymik2020enhanced, Barredo2016an, Barredo2018synthetic}. 
Controlled flows of excitations as well quantum transport of excitations in different tweezers networks has been analyzed~\cite{lienhard2020realization,wu2022manipulating,bornet2024enhancing,perciavalle2023controlled, han2024tuning, valenciatortora2023rydberg, Li2022coherent}. The preparation and the propagation of excitations with well defined momenta in Rydberg-atom platforms has been also considered for its interest in high-energy physics quantum simulators of mesons scattering~\cite{bennewitz2024simulating, surace2021scattering} or to study emergent phenomena in confined matter~\cite{domanti2024aharonov}. 

Most, if not all,  the methods realized so far suffer of bottlenecks, ultimately limiting the quality of the  current states that can be physically achieved in the circuits. 
In particular,  current states in networks as simple as a single triangle  have been realized \cite{lienhard2020realization} 
but  states of Rydberg excitations propagating in more generic circuits remain  challenging. On the other hand, methods  for achieving current states in ring-shaped networks  have been proposed, but the results are limited by currents that can decay on specific time scale    \cite{perciavalle2023controlled}. 

%%%%%%%%%%%%%%%%%%%%%%%%%%%%%%%%%
\begin{figure}[!b]
\centering
\includegraphics[width=1\columnwidth]{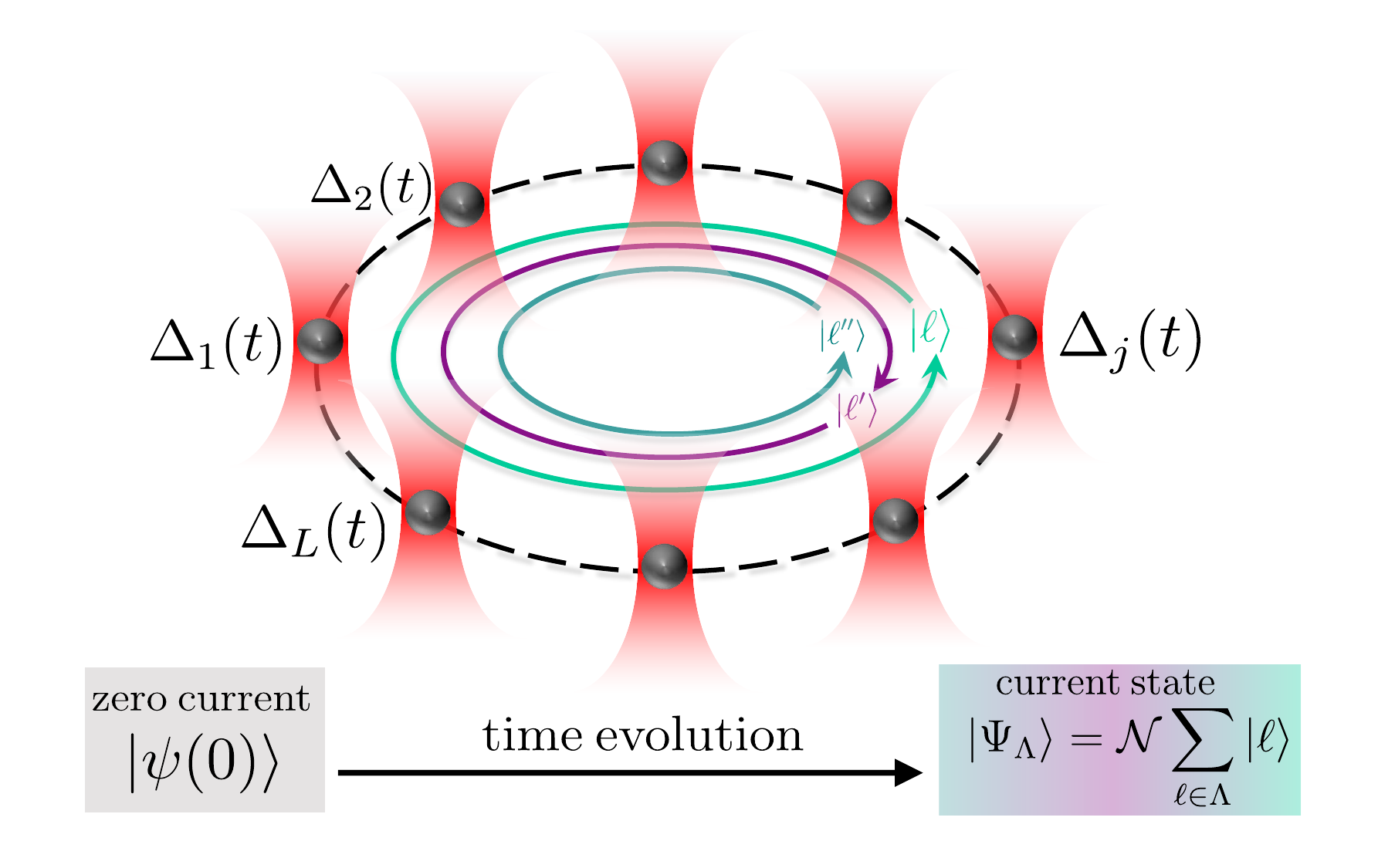}
\caption{Scheme to generate quantum superpositions of current states. An array of Rydberg atoms arranged in a ring-shaped geometry is initialized in a zero-current state $\ket{\psi(0)}$ with a single localized excitation. The system is then let evolve in time under the Hamiltonian \eqref{eq:tot_Ham}, in which the local detunings $\Delta_j(t)$ are extracted from a QOC procedure. The target state is a generic superposition $\ket{\Psi_{\Lambda}}$ of quantum states with winding numbers $\ell$, where $\Lambda$ is the set of winding numbers participating to the superposition.
Once the state is reached, the detunings are switched off and the total current remains persistent.}
\label{fig:Sketch}
\end{figure}
%%%%%%%%%%%%%%%%%%%%%%%%%%%%%%%%%

In this paper, we present a method to engineer stable current states  with arbitrary winding numbers in a ring-shaped tweezer network. Our approach is based on a suitable engineering of quantum states through optimization algorithms and quantum optimal control (QOC) protocols~\cite{Koch2022quantum, Werschnik2007quantum, brif2010control,ansel2024introduction, glaser2015training, muller2022one,zhao2023fast,pagano2024optimal,li2024generation}. 
We demonstrate that, by our approach, we can   engineer not only states with a single winding number but also  superposition of quantum current states characterized by a larger set of winding numbers. Remarkably, we shall see that such states  can be achieved as eigenstates of the Hamiltonian in the single-excitation sector; as such the target states  lead to a persistent flow, beyond the aforementioned limitations.  Specifically, we employ a gradient ascent pulse engineering (GRAPE) algorithm~\cite{Khaneja2005optimal, goerz2022quantum}, which is a gradient-decendent-based strategy to maximize the overlap between a target state $\ket{\psi_{\rm targ}}$ and a time-evolved state $\ket{\psi(t)}$ by tuning the local control parameters of the system. The target state is obtained by discretizing the time domain, which means arranging the control parameters in pulses of fixed duration. In our case, the control parameters are local detunings whose time shape are arranged in such a way to obtain a target state that is as close as possible to a state with a well defined current. We devote special attention to quantum superpositions of current states. We show how the winding numbers participating to the superposition can be used to control the chirality and the nature of the flow. The adopted protocol is sketched in Fig.~\ref{fig:Sketch}.

The paper is structured as follows.
In Sec.~\ref{sec:model} we introduce the Hamiltonian describing the physical system, the target state we want to engineer and the protocol we utilize. We then study the possibility to engineer both single current states (Sec.~\ref{sec:sing_curr}) and superpositions of different current states (Sec.~\ref{sec:two_totcurr}), paying attention to the carried total current, the detuning pattern needed to reach it, and the role of the system size. Section~\ref{sec:chiral} focuses on the dynamics of the superposition state after the optimization protocol: we observe how this is characterized by a chiral flow of excitations that can be controlled through the number and the nature of states participating to the superposition. Finally, in Sec.~\ref{sec:concl} we draw our conclusions. The Appendices report more detailed computations and simulations supporting all the findings described in the main text.

\section{Model and Methods}
\label{sec:model}

We consider an array of $L$ atoms, each of them with two dipole-coupled Rydberg states $\ket{\downarrow}_j$ and $\ket{\uparrow}_j$,
arranged in a ring-shaped configuration as in Fig.~\ref{fig:Sketch}. The dipole-dipole interaction between different Rydberg atoms in the subspace spanned by the aforementioned two levels will result in the Hamiltonian~\cite{Browaeys2020many}:
\begin{equation}
\hat{\mathcal{H}}_0 = \sum_{k\neq j}\dfrac{C_3}{d_{kj}^{3}} \big( \hat{\sigma}_{k}^{x}\hat{\sigma}_{j}^{x} + \hat{\sigma}_{k}^{y}\hat{\sigma}_{j}^{y} \big) ,
\label{eq:int}
\end{equation}
where $\hat \sigma^\alpha_j$ are the spin-1/2 Pauli matrices ($\alpha=x,y,z$) coupling the Rydberg states of the $j$th atom, $C_3$ is the interaction strength contribution given by the dipole matrix elements~\cite{barredo2015coherent,Browaeys2020many}, $d_{kj}=2R\sin(\pi|j-k|/L)$ is the distance between two atoms located at positions $j$ and $k$, $R$ being the ring radius ($\hbar$ is fixed to $1$). The interaction is isotropic, since we are supposing that the quantization axis is orthogonal with respect to the plane in which atoms are located~\cite{Chen2023continuous}.

Our goal is to realize single current quantum states and their superpositions. A single current state reads
\begin{equation}
    \ket{\ell}=\dfrac{1}{\sqrt{L}}\sum_{j=1}^L e^{i2\pi\ell j / L}\hat{\sigma}_j^+ \ket{\downarrow,\ldots,\downarrow} ,
    \label{eq:currstate}
\end{equation}
where $\ell=1, \ldots, L$ denotes the winding number and $\hat \sigma^\pm_j = \tfrac12 (\hat \sigma^x_j \pm i \hat \sigma^y_j)$. The expectation value of the nearest-neighbor current operator \begin{equation}
    \hat{\mathcal{I}} = -i \frac{J_{n.n.}}{L} \sum_{j=1}^L \big( \hat{\sigma}_{j}^+\hat{\sigma}^-_{j+1} - \rm H.c. \big)
    \label{eq:current_op}
\end{equation}  
on the $\ket{\ell}$ state is
\begin{equation}
\label{eq:current}
\braket{\ell|\hat{\mathcal{I}}|\ell} =
%&= -\frac{i {L}\sum_{j=1}^{L}J_{n.n.}\braket{\ell|\hat{\sigma}_{j}^+\hat{\sigma}^-_{j+1} - \rm H.c.|\ell}=\\& = 
2 \frac{J_{n.n.}}{L} \sin (\frac{2\pi\ell}{L}),
\end{equation}
$J_{n.n.}=C_3/(2R\sin(\pi/L))^3$ being the nearest-neighbor hopping strength. To reach a target state as the one in Eq.~\eqref{eq:currstate}, we initialize the system in a single localized excitation state with zero current, $\ket{\psi(0)} = \ket{\uparrow,\downarrow,\ldots,\downarrow}$, and then apply the Hamiltonian
\begin{equation}
    \hat{\mathcal{H}}(t) = \hat{\mathcal{H}}_0 +\sum_{j=1}^L \Delta_j(t) \, \hat{\sigma}_j^z
    \label{eq:tot_Ham},
\end{equation}
where the control parameters $\Delta_j(t)$ are local time-dependent detunings properly adjusted to achieve the desired state $\ket{\psi_{\rm targ}}$ at a certain target time $T_{\rm targ}$. The localized detunings are experimentally realizable through the coupling via an addressing beam of a low-lying state with the $\ket{\uparrow}$ state; the low-lying state is chosen to be dipole coupled with $\ket{\uparrow}$ but not with $\ket{\downarrow}$~\cite{bornet2024enhancing,de2017optical}. While the energy separation between two Rydberg states is on the microwave scale, the separation between the Rydberg states and the low-lying state is of the order of $1\mu$m in terms of wavelength. 
If the addressing beam is highly detuned, the $\ket{\uparrow}$ state experiences an effective shift $\Delta_j$ of the order of few MHz controlled through frequency and power of the addressing beam~\cite{bornet2024enhancing,de2017optical}.  Due to the coupling with a low-lying state, the Rydberg state $\ket{\uparrow}$ can have a finite life-time which is typically on the order of hundreds $\mu$s for effective detunings of the order of few MHz. Therefore, for dynamics on the order of few $\mu$s, it can be negligible~\cite{bornet2024enhancing,de2017optical}.

%%%%%%%%%%%%%%%%%%%%%%%%%%%%%%%%%
\begin{figure*}[!t]
\centering
\includegraphics[width=\textwidth]{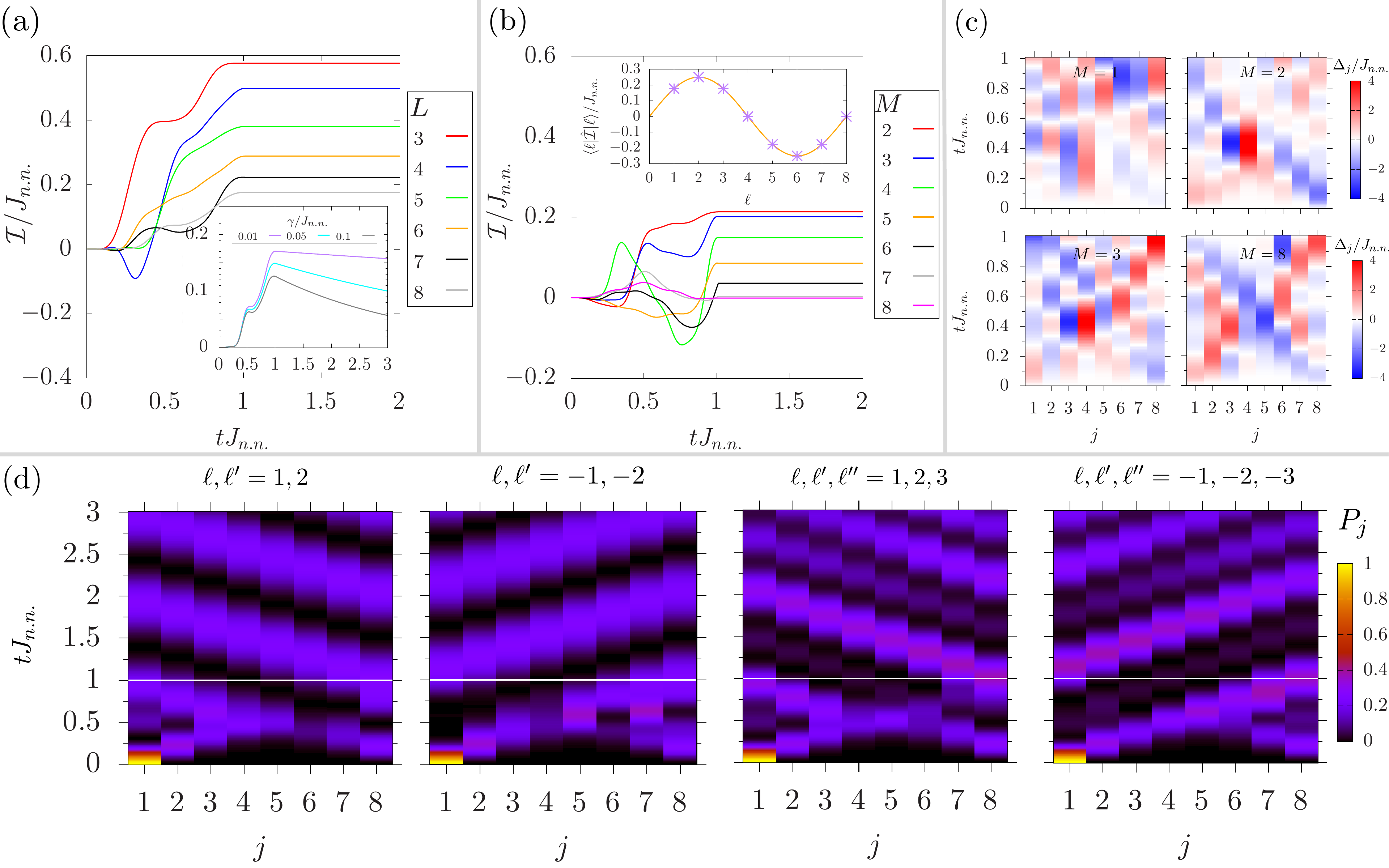}
\caption{%Dynamics of the system. 
Dynamics of the current $\mathcal{I}(t)=\braket{\psi(t)|\hat{\mathcal{I}}|\psi(t)}$ when the target state is: (a) a single current state $\ket{\ell=1}$, for different system sizes; the inset in (a) reports the behavior of the current in the presence of pure dephasing before and after the optimization protocol for different values of the dephasing rate $\gamma$, for $L=8$ (more details are reported in \color{blue}App.~\ref{app:rob}\color{black}). \color{blue}(b)\color{black}: a superposition $\ket{\Psi^{(M)}} \propto \sum_{\ell=1}^M\ket{\ell}$, for different values of $M$ (with $L=8$); the inset in (b) reports the expectation value $\braket{\ell|\hat{\mathcal{I}}|\ell}/J_{n.n.}$, where purple points are integer values of $\ell$ [cf.~Eq.~\eqref{eq:current}].
(c): Detuning profiles in time and space used to realize the target state $\ket{\Psi^{(M)}}$; results are reported for $L=8$ and for different $M$. (d): Local population dynamics for various target superposition states: $\ket{1}+\ket{2}$, $\, \ket{-1}+\ket{-2}$, \, $\ket{1}+\ket{2}+\ket{3}$, $\, \ket{-1}+\ket{-2}+\ket{-3}$; the white line denotes the target time $T_{\rm targ}=J_{n.n.}^{-1}$. In all panels $T_{\rm targ}=J_{n.n.}^{-1}$ and $\Delta t = T_{\rm targ}/100$.}
\label{fig:superp}
\end{figure*}
%%%%%%%%%%%%%%%%%%%%%%%%%%%%%%%%%

The optimization protocol is performed through the GRAPE algorithm~\cite{Khaneja2005optimal,goerz2022quantum}, based on the discretization of the time interval and maximization of the fidelity between the target $\ket{\psi_{\rm targ}}$ and the time-evolved $\ket{\psi(t)}$ states through a gradient descent optimization protocol~\cite{ruder2016an}. Physically, this is achieved by laser pulses of fixed duration $\Delta t$ that bring to the wanted effective inhomogeneous detuning profile $\Delta_{j}(t)$. Once the target state is reached, we switch off the detuning and let the system evolve under the bare Hamiltonian $\hat{\mathcal{H}}_0$ of Eq.~\eqref{eq:int}. The crucial point is that the states $\ket{\ell}$ with a single excitation are eigenstates of both the bare Hamiltonian and the current (see App.~\ref{app:A}),  therefore, after the target time and in absence of decoherence, they do not evolve and the current is conserved. Thus the system is able to support a persistent current. In contrast, a superposition of non-degenerate current states evolves in time. Since the population distribution of such superposition is not homogeneous, the dynamics will result in a directional flow of the latter,  controllable through the choice of the $\ell$ winding numbers participating to the superposition. The possible presence of decoherence causes a decay in time of the current.

In this specific work, we focus on the superposition of two and three current states. The dynamics under $\hat{\mathcal{H}}_0$ of the superposition of two current states is guided by a relative phase between them, which generates the directional flow of excitations. The superposition of three current states gives rise to a dynamics characterized by a more complex flow of excitations. To monitor such properties, we analyze the expectation value of local populations, currents, and correlation functions. Remarkably, this kind of measurements can be experimentally accessed in Rydberg-atom platforms~\cite{bornet2024enhancing,Chen2023continuous,chen2023spectroscopy}, making them
an ideal candidate to characterize these type of states.
We note that, in our work, the measurement of observables is not used for the optimization protocol but as a read-out of the quantum dynamics we observe.

\section{Single current state}
\label{sec:sing_curr}
%{\it Single current state.} --- 
We first focus on the realization of the state in Eq.~\eqref{eq:currstate}. 
We choose a target time $T_{\rm targ}=J_{n.n.}^{-1}$ and time spacing $\Delta t = T_{\rm targ}/100$. Using as reference the realistic value of the nearest-neighbor hopping $J_{n.n.}\sim 1$MHz, the duration $\Delta t$ of each pulse is of the order of $10$ns, which is within experimental reach~\cite{bornet2024enhancing}. We extract the values of $\Delta_j(t)$ needed to reach $\ket{\psi_{\rm targ}} = \ket{\ell}$ through the GRAPE algorithm. 
Fig.~\ref{fig:superp}(a) reports the time behavior of the current operator expectation value when the target state is the single current state $\ket{\ell=1}$. In the time interval $[0,T_{\rm targ}+\Delta t]$ the state evolves under the Hamiltonian~\eqref{eq:tot_Ham}, while at $t\geq T_{\rm targ}+\Delta t$ the control parameters are switched off and the evolution continues under the effect of the dipolar bare Hamiltonian~\eqref{eq:int}, with the current remaining stable and persistent for all system sizes considered under unitary evolution.  $T_{\rm targ}=J_{n.n.}^{-1}$ is sufficiently big to reach the target states with high fidelity using $100$ pulses. A more detailed analysis of the optimal target time and $\Delta t$ is performed in App.~\ref{app:rob}.

We remark that the current is originated by the coherent nature of the state, therefore the possible presence of decoherence causes its decay in time: after the target time it decays as $\exp(-4\gamma t)$. Numerical simulations are reported in the inset of Fig.~\ref{fig:superp}(a) and, more extensively, in App.~\ref{app:rob} The current remains nonzero for time-scales larger than $J_{n.n.}^{-1}$ if the dephasing rate is smaller than $J_{n.n.}$. Another source of errors is the imperfect realization of the desired detuning pattern: in App.~\ref{app:rob} we study how the protocol is affected by noise in the detuning and show that, for a noise smaller than or comparable to $J_{n.n.}$, the protocol remains robust.

\section{Total current of the superposition state}
\label{sec:two_totcurr}
%{\it Superposition of current states.} --- 

As a second step, we focus on the realization of more exotic states as
\begin{equation}
\ket{\Psi_\Lambda}=\mathcal{N}\sum_{\ell\in \Lambda}\ket{\ell},
\label{eq:currentSuperpos}
\end{equation}
$\Lambda$ being the set of winding numbers that participate to the superposition and $\mathcal{N}$ a normalization constant. We consider the matrix element
\begin{equation}
\braket{\ell|\hat{\mathcal{I}}|\ell '}=-i\frac{J_{n.n.}}{L}\sum_{j=1}^L \braket{\ell|\hat{\sigma}_{j}^+\hat{\sigma}^-_{f(j+1)} - \rm H.c.|\ell '},
\end{equation}
where $f(j\!+\!1)=j\!+\!1$ for $j\neq L$, while $f(L\!+\!1)=1$. Using the explicit form of the current state, we obtain
\begin{equation}
\braket{\ell|\hat{\mathcal{I}}|\ell '}\propto \big( e^{i \frac{2\pi}{L} \ell'} - e^{-i \frac{2\pi}{L} \ell}\big) \, e^{-\frac{i2\pi(\ell - \ell ')}{L}}\dfrac{1-e^{i 2\pi (\ell - \ell ')}}{1-e^{i \frac{2\pi}{L} (\ell - \ell ')}},
\label{eq:mat_element}
\end{equation}
where we consider $\ell$ and $\ell '$ integers, with $\ell - \ell ' \neq Ln$ ($n = 0,\pm 1, \pm 2, \ldots$). Under these conditions, the difference $\ell - \ell '$ is an integer number, although different from zero and from multiples of $L$. The numerator of Eq.~\eqref{eq:mat_element} is zero, while the denominator is different from zero, meaning that $\braket{\ell|\hat{\mathcal{I}}|\ell '}=0$. In contrast, in the case $\ell - \ell ' = Ln$, the formula~\eqref{eq:mat_element} does not apply: since the states $\ket{\ell}$ and $\ket{\ell ' = \ell - Ln}$ coincide, the matrix element is $\braket{\ell|\hat{\mathcal{I}}|\ell - Ln}=\braket{\ell|\hat{\mathcal{I}}|\ell}$, which coincides with Eq.~\eqref{eq:current}. Thus, the total current carried by $\ket{\Psi_{\Lambda}}$ is the weighted sum of the currents carried by each superposed state
\begin{equation}
\braket{\Psi_{\Lambda}|\hat{\mathcal{I}}|\Psi_{\Lambda}}=\mathcal{N}^2\sum_{\ell,\ell ' \in \Lambda}\braket{\ell|\hat{\mathcal{I}}|\ell '}=\mathcal{N}^2\sum_{\ell \in \Lambda }\braket{\ell|\hat{\mathcal{I}}|\ell}.
\end{equation}
Considering a superposition of contiguous winding numbers, that is, $\Lambda=\{1,\ldots,M\}$:
\begin{equation}
   \ket{\Psi_{\Lambda}} \equiv \ket{\Psi^{(M)}}=\dfrac{1}{\sqrt{M}}\sum_{\ell=1}^M\ket{\ell},
\end{equation}
its total current is
\begin{equation}
\braket{\Psi^{(M)}|\hat{\mathcal{I}}|\Psi^{(M)}}=\dfrac{1}{M}\sum_{\ell = 1}^M\dfrac{2J_{n.n.}}{L}\sin \left(\dfrac{2\pi\ell}{L} \right)
\end{equation}
and can be written in the closed form
\begin{equation}
\braket{\Psi^{(M)}|\hat{\mathcal{I}}|\Psi^{(M)}} = \dfrac{2J_{n.n.}}{ML} \, \dfrac{ \sin(\pi M/L) \, \sin \big[ \pi (M+1)/L \big]}{\sin(\pi/L)}.
\end{equation}

%\subsection{Superposition of $M$ current states}
%
The behavior in time of the current under a QOC protocol targeted to reach $\ket{\psi_{\rm targ}} = \ket{\Psi^{(M)}}$ is reported in Fig.~\ref{fig:superp}(b), until and after the target time, for different values of $M$.  In all the considered cases, the time behavior of the current is irregular until the target state is reached. In contrast, after $T_{\rm targ}$, the system continues to evolve under the bare Hamiltonian and the current is conserved (the behavior of $\braket{\ell|\hat{\mathcal{I}}|\ell}$ is reported in the inset as a reference). For $M=L$ and $M=L-1$ the total current of the target state has an equal contribution from negative and positive current, and thus it is zero overall. Other target states have a small but nonzero current. Decreasing $M$ from $L-2$ to $L/2$, we reduce the number of negative current states that contribute to the total current, thus the overall value increases. For $M<L/2$, the current keeps increasing by further decreasing $M$, due to the change of the weight given from different normalizations of the state. As in the case of single current states, the presence of decoherence causes a decay in time of the current after the target time and the protocol is robust in presence of noise in the detuning (see App.~\ref{app:rob}).

Detuning profiles needed to realize different current states are reported in Fig.~\ref{fig:superp}(c). We observe that they are characterized by small frequency oscillations in time, with a rather complex behavior that strongly depends on the target state. Concerning their magnitude, if we consider the reference value $J_{n.n.}\sim 1$MHz, the required values of detuning are of the order of few MHz, which are experimentally feasible~\cite{de2017optical}. In App.~\ref{app:B} we completed the detuning analysis studying their shape for different $\Delta t$, fixed $T_{\rm targ}$. Increasing $\Delta t$ the general shape of the optimal pulses does not change, while it becomes more discretized.

\section{Chiral flows of excitations}
\label{sec:chiral}
%{\it Chiral flows of excitations.} --- 

Unlike single current states, a superposition state presents an inhomogenous distribution of the population, which can be computed using the explicit form of the superposition of $M$ current states
\begin{equation}
\ket{\Psi^{(M)}} = \dfrac{1}{\sqrt{ML}}\sum_{j=1}^L \sum_{\ell=1}^M e^{i 2\pi\ell j / L}\hat{\sigma}_j^+\ket{0},
\end{equation}
where we defined $\ket{0}=\ket{\downarrow,\ldots,\downarrow}$.
For $j=L$, it is straightforward to obtain the population distribution $P_L^{(M)}=\big\langle \ket{j\!=\!L}\bra{j\!=\!L} \big\rangle = M/L$, for $j\neq L$ we obtain
\begin{equation}
    P_{j\neq L}^{(M)} = \braket{\ket{j}\bra{j}}=\dfrac{1}{ML}\dfrac{\sin^2\left(M\pi j / L\right)}{\sin^2\left(\pi j / L\right)},
\end{equation}
where $\ket{j}\equiv\hat{\sigma}_j^+\ket{0}$.
Due to interference between the various superimposed phases, the population distribution is not homogeneous. Nonetheless, the total current is conserved, thus we expect the generation of a controlled directional flow of excitations after the optimization protocol.  

\subsection{Population dynamics of a superposition state}
We now consider the time-evolution of the state $\ket{\Psi^{(M)}}$ under the bare Hamiltonian $\hat{\mathcal{H}}_0$: 
\begin{align}
    \ket{\Psi^{(M)}(t)}&=\dfrac{1}{\sqrt{M}}\sum_{\ell=1}^M e^{-iE_{\ell} t}\ket{\ell}, %\nonumber \\ &=\dfrac{1}{\sqrt{ML}}\sum_{j=1}^L \sum_{\ell=1}^L e^{i2\pi\ell j / L}e^{-iE_{\ell}t}\hat{\sigma}_j^+\ket{0}, 
\end{align}
where $E_\ell$ is the eigenvalue associated with $\ket{\ell}$.
For a superposition of two states $\ket{\ell}$ and $\ket{\ell '}$, the population distribution is given by
\begin{equation}
P_j^{(2)}(t) =\dfrac{1}{L}\left\{ 1 + \cos\left[ \dfrac{2\pi (\ell' - \ell) j}{L} - \omega_{\ell\ell '} t \right] \right\}
\label{eq:pop_distr}
\end{equation}
where
\begin{equation}
\omega_{\ell \ell '}=E_{\ell }- E_{\ell '} = 4 \sum_{k}\phantom{}^{'} \! J_k \Big[ \cos \big( 2\pi\ell k/L \big) - \cos \big( 2\pi\ell' k/L \big) \Big],
\end{equation}
with $\sum_{k}\phantom{}^{'}$ accounting for the parity of $L$ and being defined in App.~\ref{app:A}. The time-dynamics of the two current superposition state is given by 
\begin{equation}
   \ket{\Psi^{(2)}(t)}= e^{-i\hat{\mathcal{H}}_0t}\ket{\Psi^{(2)}} = \tfrac{1}{\sqrt{2}}[\ket{\ell}+e^{i\omega_{\ell \ell '}t}\ket{\ell '}].
   \label{eq:ellellp_evo}
\end{equation}
Considering the specific state 
\begin{equation}
  \ket{\Psi^{(2)}}=\tfrac{1}{\sqrt{2}}[\ket{\ell=1}+\ket{\ell '=2}],
  \label{eq:sup12}
\end{equation}
its population distribution \eqref{eq:pop_distr} is peaked at $j=L$. During the time-evolution, the two superposed current states acquire a relative phase $\phi(t)=\omega_{\ell \ell '}t$. 

We parametrize the relative dynamical phase introducing $t_n=2\pi n / L\omega_{\ell\ell '}$. The time-evolved state $\ket{\Psi^{(2)}(t_n)}$ populations distribution
\begin{equation}
P_j^{(2)}(t_n) = \tfrac{1}{L} \Big\{ 1 + \cos\left[2\pi(j+n)/L \right] \Big\} 
\end{equation}
are peaked on $j=L-n$. Thus, for a system initialized in $\ket{\Psi^{(2)}}$, $\ket{\Psi^{(2)}(t_n)}$ is the state whose distribution peak is translated by $n$ sites in anti-clockwise direction, for positive $n$. The shape of the distribution is unchanged, an excitation blob moves rigidly in the ring. In a time $\tau_{n-\rm{transl}}=2\pi n / L\omega_{\ell \ell '}$, the peak moves of $n$ sites in anti-clockwise/clockwise direction. In $\tau_{(L/2)-\rm{transl}}=\pi /\omega_{\ell \ell '}$ the state $\ket{\psi_{\perp}}= \tfrac{1}{\sqrt{2}} ( \ket{\ell=1}-\ket{\ell '=2} )$ is reached, its population distribution peak is in the diametrically opposite position of those of $\ket{\Psi^{(2)}}$. The states $\ket{\Psi^{(2)}}$ and $\ket{\psi_{\perp}}$ are orthogonal. $\tau_{(L/2)-\rm{transl}}$ coincides with the quantum speed limit (QSL) according to the Mandelstam-Tamm (MT) derivation~\cite{mandelstam1945uncertainty,deffner2017quantum, giovannetti2003quantumlimits}. More precisely, the QSL is the minimal time required to evolve between two orthogonal states under a time-independent Hamiltonian (see App.~\ref{app:QSL}). Given a superposition of two current states $\ell$ and $\ell '$, which are eigenstates of the dipolar XY Hamiltonian $\hat{\mathcal{H}}_0$, the MT-QSL is
\begin{equation}
\tau_{\rm MT}=\dfrac{\pi}{\omega_{\ell\ell '}},
\qquad 
\Delta E_0^2 \equiv \braket{\hat{\mathcal{H}}_0^2} - \langle \hat{\mathcal{H}}_0 \rangle^2 = \frac{\omega_{\ell \ell '}^2}{4},
\end{equation}
which implies $\tau_{\rm MT}=\tau_{(L/2)-\rm transl}$. The distribution rotates in the ring, its peak moves saturating the MT-QSL. The velocity with which the excitation blob is moving is $v_{\ell\ell '}=a/\tau_{1-\rm{transl}}=La\omega_{\ell\ell '}/2\pi$, $a$ being the nearest-neighbor distance.

\subsection{Population under the optimization protocol}

We analyze the dynamics of the local populations for different target states $\ket{\Psi_\Lambda}$: Fig.~\ref{fig:superp}(d) reports the population dynamics of the superposition of two and three current states, each of them taken positive and negative, so for each state we take its chiral counterpart. We observe that the dynamics before the target state depends on the sign of the current associated to target state; two current states with opposite current have the same population distribution but different phase pattern, therefore the QOC protocol strongly depends on the target. 
Once the target state is reached, a directional flow of excitations is observed. 

%%%%%%%%%%%%%%%%%%%%%%%%%%%%%%%%%
\begin{figure}[!t]
\hfill
\subfigure[\empty]{\includegraphics[width=0.505\columnwidth]{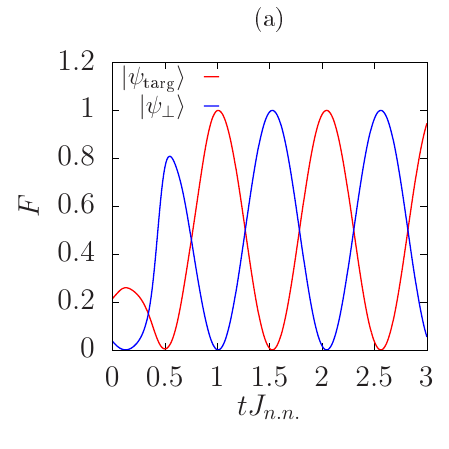}}
\hfill
\subfigure[\empty]{\includegraphics[width=0.475\columnwidth]{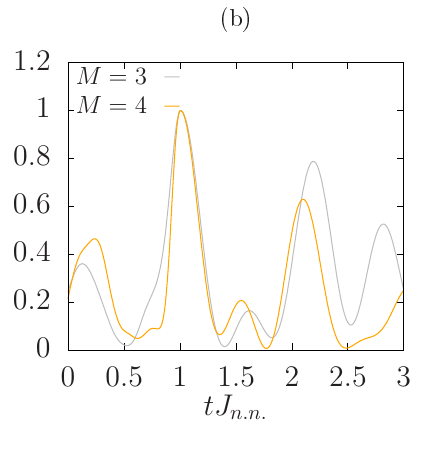}}
\hfill
\caption{(a): The fidelity between the time evolved state and either $\ket{\psi_{\rm targ}}= \tfrac{1}{\sqrt{2}}[\ket{\ell=1}+\ket{\ell '=2}]$ (red) or its orthogonal $\ket{\psi_{\perp}}=\tfrac{1}{\sqrt{2}}[\ket{\ell=1}-\ket{\ell '=2}]$ (blue). (b): The fidelity between the time evolved state and $\ket{\psi_{\rm targ}}=\tfrac{1}{\sqrt{M}} \sum_{\ell=1}^M\ket{\ell}$, for two different values of $M$. We fix $L=8$, $T_{\rm targ}=J_{n.n.}^{-1}$, and $\Delta t = T_{\rm targ}/100$.}
\label{fig:Fid_sup_M}
\end{figure}
%%%%%%%%%%%%%%%%%%%%%%%%%%%%%%%%%

A superposition of two current states presents a regular flow of excitations in which a blob of excitations centered around $j=L$ and distributed according to~\eqref{eq:pop_distr} rotates in the ring. After $\ket{\psi_{\rm targ}}$ is reached, the state evolves as~\eqref{eq:ellellp_evo}, thus showing Rabi-like oscillations between $(\ket{\ell}+\ket{\ell '})/\sqrt{2}$ and $(\ket{\ell}-\ket{\ell '})/\sqrt{2}$. We remark that the population distribution of a target state composed by two current states can be written as \eqref{eq:pop_distr}, thus, increasing the value of $\ell ' - \ell$, it is possible to increase the number of peaks and nodes of the distribution, generating a bigger number of excitation blobs that will move directionally in the ring. The analysis for $\ell \neq 0$ and $\ell'=0$ is reported in Sec.~\ref{sec:0ell}.

In the case of three current states, the time-evolved state is
\begin{equation}
\ket{\Psi^{(3)}(t)}= \tfrac{1}{\sqrt{3}} \big( \ket{\ell}+e^{i\omega_{\ell\ell '}t}\ket{\ell '} +e^{i\omega_{\ell \ell ''}t}\ket{\ell ''} \big)
\end{equation}
and the competition of two phases results in a less clean flow of exictations. However, the target state is more localized in $j=L$, thus the excitation blob that moves in the ring is narrower and higher.
Regarding the chirality, in both cases we clearly observe it, with directionality that switches by changing the sign of the winding numbers contained in the superposition; this means that the QOC works well and a chiral target state is reached.

Figure~\ref{fig:Fid_sup_M}(a) shows the fidelity between the time evolved state and the states $\ket{\psi_{\rm targ}} = \tfrac{1}{\sqrt{2}} (\ket{\ell=1}+\ket{\ell '=2})$ and $\ket{\psi_{\perp}} = \frac{1}{\sqrt{2}} (\ket{\ell=1}-\ket{\ell '=2})$, respectively computed as
\begin{equation}
F(t)=|\!\braket{\psi(t)|\psi_{\rm targ}}\!|^2 \quad \mbox{and} \quad F(t) = |\!\braket{\psi(t)|\psi_{\perp}}\!|^2.
\end{equation}
At $T_{\rm targ}=J_{n.n.}^{-1}$ the target state is reached with high fidelity, then the system evolves showing Rabi oscillations between $\ket{\psi_{\rm targ}}$ and $\ket{\psi_{\perp}}$. The excitation blob is initially localized in $j=L$ then it moves following the direction given by its total current and reaches $j=L/2$, the total current is conserved and so the flow continues and persists at any time. 

%%%%%%%%%%%%%%%%%%%%%%%%%%%%%%%%%
\begin{figure}[!t]
\centering
\includegraphics[width=\columnwidth]{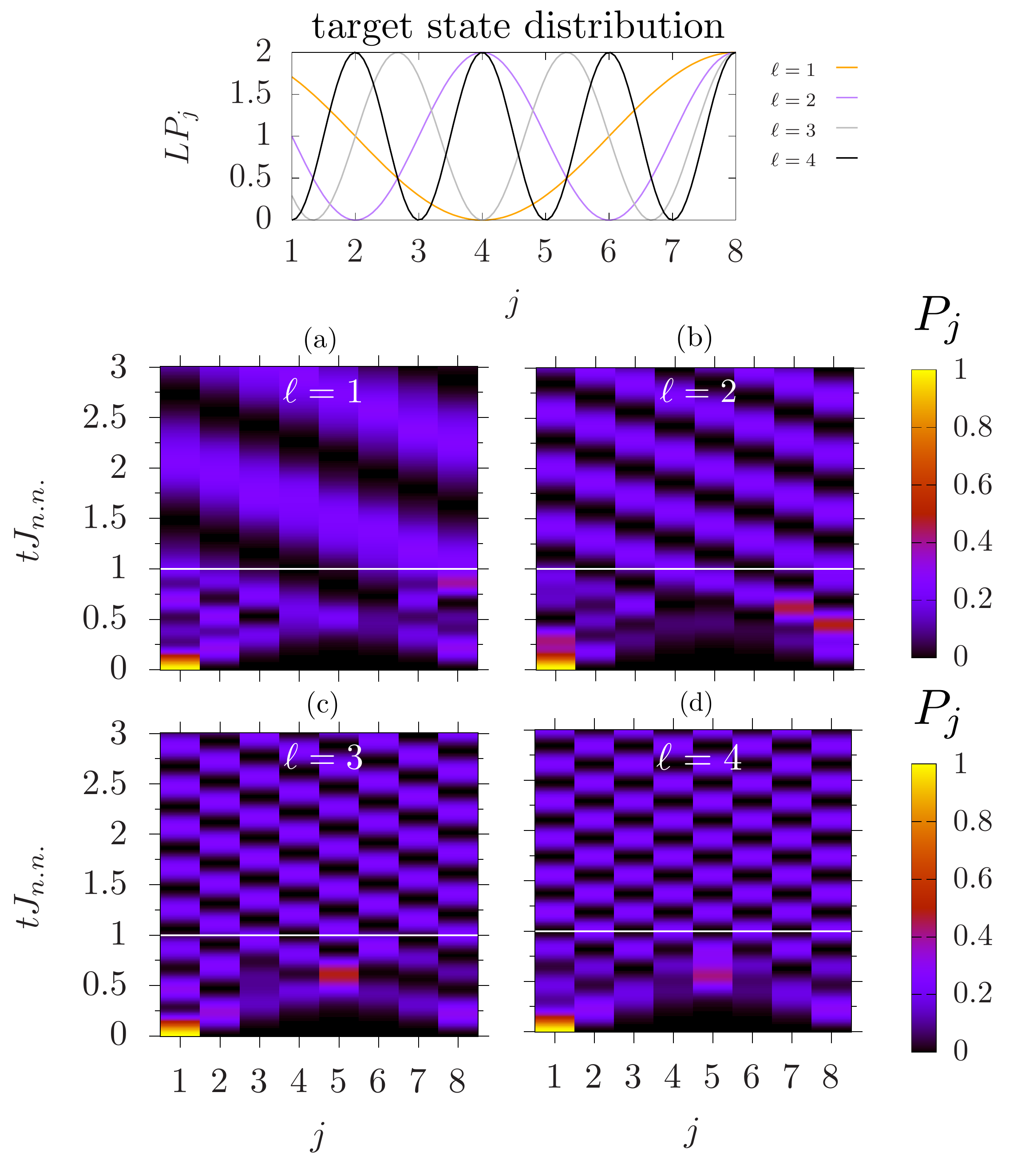}
\caption{The population dynamics for a target state $\ket{\psi_{\rm targ}} = \tfrac{1}{\sqrt{2}} \left[ \ket{\ell \neq 0} + \ket{\ell ' =0}\right]$, and for different values of $\ell$. The top panel reports the ideal distribution of the target state. (a),(b),(c),(d) report the population dynamics for $\ell=1,2,3,4$ respectively. We fix $L=8$, $T_{\rm targ}=J_{n.n.}^{-1}$, and $\Delta t = T_{\rm targ}/100$.
\color{black}}
\label{fig:superpell0}
\end{figure}
%%%%%%%%%%%%%%%%%%%%%%%%%%%%%%%%%

In the presence of more than two current states, 
multiple relative phases between current states emerge, the dynamics is can no longer be described through Rabi oscillations between two orthogonal states. Fig.~\ref{fig:Fid_sup_M}(b) reports the fidelity between the time evolved state and the superposition of three and four current state. The target state is reached with high fidelity; once the detunings are switched-off, the dynamics is not regularly oscillating, meaning that the excitation flow is complex.

\subsection{Superposition of zero and nonzero current states}
\label{sec:0ell}

We study the dynamics in the case of the target state
\begin{equation}
    \ket{\Psi_{\Lambda}}=\tfrac{1}{\sqrt{2}}\left[ \ket{\ell \neq 0} + \ket{\ell ' =0}\right],
    \label{eq:ell0}
\end{equation}
which is the superposition of two states, one with with zero and one with non zero current. The total current of the state $\mathcal{I}= \tfrac12 \braket{\ell|\hat{\mathcal{I}}|\ell}$ is half of those of the single winding state $\ket{\ell}$. The superposition is responsible for inhomogeneous local current and population distribution pattern, which causes a dynamics strongly different from those of the single current state.

for which the population distribution is given by
\begin{equation}
    P_j =\tfrac{1}{L} \big[ 1 + \cos \big(2\pi\ell j/L \big) \big],
\end{equation}
which is a simplified version of~\eqref{eq:pop_distr}. Increasing the value of $\ell$, the number of peaks of the distribution increases. Thus, $\ell$ can be used to control to number of blobs that directionally move in the ring under a total current $\mathcal{I}=\tfrac12 \braket{\ell|\hat{\mathcal{I}}|\ell}$. Figure~\ref{fig:superpell0} shows the population dynamics for different values of the winding $\ell$ of the target state. 
From the target state distribution reported in the top panel of Fig.~\ref{fig:superpell0}, we observe that increasing the value of $\ell$ the number of peaks of the population distribution increases, while its width decreases. The bottom panels (a),(b),(c),(d) show the population dynamics for increasing values of $\ell$. The possibility to control the number of peaks that move in the ring with velocity $v_{0\ell}$ constitute an important resource to track the nature of the superposition state directly from measurement of the population distribution dynamics.

\section{Final remarks}
\label{sec:concl}

We demonstrated how quantum optimal control techniques are suitable candidates to realize persistent current states in dipolar ring-shaped Rydberg atom networks, through time-dependent arranging of local detunings.
Single current states can be realized with high fidelity if the duration of the detuning pulses remains reasonably small with respect to the total target time. Once the detuning are switched-off, the current is conserved in time. 
Single current states can be realized with high fidelity if the duration of the detuning pulses remains reasonably small with respect to the total target time. Once the detunings are switched-off, the current is conserved in time. As such, our results allow to make progress on the current states achieved through the  Laguerre-Gauss protocol proposed in \cite{perciavalle2023controlled}.
More exotic superpositions of current states can be realized as well. Such states are  characterized by an inhomogeneous population distribution and nonzero current resulting in a directional flow of excitations in the ring that ideally persists at any time. Superposition represents an improvement in terms of controllability of the flow,
its nature and its velocity can be carefully controlled by selecting the winding numbers that participate to the superposition.
Superposing two current states is particularly interesting, since it generates Rabi oscillations between the target state and its mirror state in the ring. The latter is reached through a clean excitation flow, whose directionality can be controlled through the sign of the winding numbers of the states participating to the superposition. 
The excitation velocity is controllable as well. The difference between the two winding numbers involved in the superposition can be used to control the number of excitation blobs that move in the ring. The system is inevitably subject to decoherence: pure dephasing causes a decay  of the current in time however, for dephasing rates smaller than the typical hopping energy, the current remains far from zero on time scales larger than $J_{n.n.}^{-1}$ (see App.~\ref{app:rob}).

We remark that based on the current experimental know-how in the field (allowing to monitor Rydberg excitations populations, correlation functions and currents through single and multi-basis measurements~\cite{bornet2024enhancing, chen2023spectroscopy, Chen2023continuous}) our current states can be realistically tracked. Moreover, the excitation density distribution in superposition current states is inhomogeneous and controllable. Its dynamics reflects the presence of a current and permits an experimental detection of it without monitoring correlations. We point out that through our approach we are able to quantify the effect of noise in the detunings on the protocol (see App.~\ref{app:rob}), which results to be robust. 

The realization of quantum currents through optimization protocols  paves the way for the study of  quantum phenomena based on the coherence of persistent currents in Rydberg-atom quantum simulators~\cite{bornet2024enhancing}.  The engineering and analysis of more complex quantum states like superposition of different winding number configurations constitute an interesting future perspective of this work. We also remark that the approach proposed here is not the only possible to address the problem, the application of different QOC protocols possibly combined with optimized numerical methods could be a natural continuation of this work~\cite{goerz2022quantum,krotov1996global,doria2011optimal, caneva2011chopped, catalano2024numerically}. Moreover, machine learning based methods in which the optimization is performed through the training of a neural network via the measurement of observables constitute a valid alternative to QOC protocols. They have been proposed in bosonic matter-wave circuits~\cite{haug2021machine} and would define an interesting venue for future research also in the context of Rydberg-atom platforms.

%Machine learning based methods, as they have been applied to bosonic matter-wave circuits~\cite{haug2021machine},  would define an interesting venue for future research.

\acknowledgements

We thank Enrico C. Domanti, Wayne J. Chetcuti, and Alessio Lerose for useful discussions. The Julian Schwinger Foundation grant JSF-18-12-0011 is acknowledged. OM also acknowledges support by the H2020 ITN ``MOQS" (grant agreement number 955479) and MUR (Ministero dell’Università e della Ricerca) through the PNRR MUR project PE0000023-NQSTI. Numerical computations have been performed using the Julia packages \texttt{Grape.jl}~\cite{goerz2023grape} within the QuantumControl Julia framework~\cite{goerz2023quantumcontrol} and \texttt{QuantumOptics.jl}~\cite{Kramer2018quantum}.

\appendix
\section{The current state is an eigenstate of the Hamiltonian}
\label{app:A}
We consider a ring, whose sites are labeled as $j=1,2,\ldots,L$, described by the generic Hamiltonian
\begin{equation}
    \hat{\mathcal{H}} = \sum_{j\neq i} J_{|j-i|}\hat{h}_{j,i}
    \label{eq:S1}
    \end{equation}
where $\hat{h}_{j,i}=\hat{h}_{i,j}$, while the coupling depends on the absolute value of the difference between the sites and $J_{|j-i|}=J_{L-|j-i|}$ due to the ring-shaped geometry. The Hamiltonian can be rewritten as
\begin{equation}
    \hat{\mathcal{H}} = \sum_{k}\phantom{}^{'} \sum_{j=1}^{L} J_{|k|} \hat{h}_{j,f(j+k)},
\label{eq:S2}
\end{equation}
where $f(j + k)$ takes into account the ring geometry of the system
and is defined as
\begin{align}
& f(j+k)=
    \begin{cases}
        j+k \qquad \quad \; \operatorname{if} \; j+k \leq L, \\
        (j+k)-L \, \: \: \operatorname{if} \; j+k > L,
    \end{cases}
    \!\!(k>0); \nonumber
\\
& f(j+k)=
    \begin{cases}
        j+k \qquad \quad \; \operatorname{if} \; j+k > 0, \\
        (j+k)+L \, \: \: \operatorname{if} \; j+k \leq 0,
    \end{cases}
    \!\!(k<0), \nonumber
\end{align}
while 
\begin{eqnarray}
    \sum_k \phantom{}^{'} & = & \sum_{k=-L/2+1,(k\neq 0)}^{L/2} \quad(\mbox{$L$ even}); \nonumber \\ 
    \sum_k \phantom{}^{'} & = & \sum_{k=-\floor{L/2},(k\neq 0)}^{\floor{L/2}} \quad\; (\mbox{$L$ odd}). \nonumber
\end{eqnarray} 
%The function $f$ is responsible for the correct labeling of the coupling between various sites in the ring. 
In fact, for two sites separated by a number $k$ of links, the value $j+k$ may exceed the labeling range $1,2,\ldots,L$ from above or below; the function $f$ establishes the correct labeling of those sites. For instance, using Eq.~\eqref{eq:S2}, for $L=3$ we get
\begin{eqnarray}
\hat{\mathcal{H}} & = & J_{|1|}\hat{h}_{1,f(0)} + J_{|1|}\hat{h}_{2,f(1)} + J_{|1|}\hat{h}_{3,f(2)} \nonumber \\
& + & J_{|1|}\hat{h}_{1,f(2)} +
J_{|1|}\hat{h}_{2,f(3)} +
J_{|1|}\hat{h}_{3,f(4)}.
\end{eqnarray}
using the definition of $f$ and $\hat{h}_{j,i}=\hat{h}_{i,j}$ we can rewrite the Hamiltonian as
\begin{equation}
\hat{\mathcal{H}} = 2J_{|1|}\left(\hat{h}_{1,2} + \hat{h}_{2,3} + \hat{h}_{1,3}\right)    
\end{equation}
which is the same as \eqref{eq:S1}. For $L=4$ and \eqref{eq:S2} we get
\begin{eqnarray}
\hat{\mathcal{H}} & = & J_{|1|}\left(\hat{h}_{1,f(0)} + \hat{h}_{2,f(1)} + \hat{h}_{3,f(2)} + \hat{h}_{4,f(3)} \right) \nonumber \\ 
&+ & J_{|1|}\left(\hat{h}_{1,f(2)} + \hat{h}_{2,f(2)} + \hat{h}_{3,f(3)} + \hat{h}_{4,f(5)} \right) \nonumber \\ & + & J_{|2|}\left(\hat{h}_{1,f(3)} + \hat{h}_{2,f(4)} + \hat{h}_{3,f(5)} + \hat{h}_{4,f(6)} \right),
\end{eqnarray}
which can be rewritten as 
\begin{equation}
\hat{\mathcal{H}} =2J_{|1|}\left(\hat{h}_{1,2} + \hat{h}_{2,3} + \hat{h}_{3,4} + \hat{h}_{4,1} \right) + 2J_{|2|}\left(\hat{h}_{1,3} + \hat{h}_{2,4} \right)  
\end{equation}
which is the same as \eqref{eq:S1}.

Let us now consider the Hamiltonian
\begin{eqnarray}
    \hat{\mathcal{H}}_k & = & \tilde{J}_k \sum_{j=1}^{L} \Big( \hat{\sigma}_j^+\hat{\sigma}_{f(j+k)}^- + {\rm H.c.} \Big) \nonumber \\
    & =& J_k \sum_{j=1}^{L} \Big( \hat{\sigma}_j^x\hat{\sigma}_{f(j+k)}^x +  \hat{\sigma}_j^y\hat{\sigma}_{f(j+k)}^y \Big);
    \label{eq:Ham_nn}
\end{eqnarray}
where $\tilde{J}_k = 2 J_k$ and separately discuss the cases $k>0$ and $k<0$.  For $k>0$, when applying the Hamiltonian~\eqref{eq:Ham_nn} on the current state $\ket{\ell}$ we obtain
\begin{equation}
\hat{\mathcal{H}}_k\ket{\ell} = \dfrac{\tilde{J}_k}{\sqrt{L}} \sum_{j,j'=1}^{L} e^{\frac{i2\pi\ell j'}{L}} \Big( \hat{\sigma}_j^+\hat{\sigma}_{f(j+k)}^- + {\rm H.c.} \Big) \hat{\sigma}_{j'}^{+}\ket{0},
\end{equation}
where $\ket{0}\equiv\ket{\downarrow,\ldots,\downarrow}$. Now, using $[\hat{\sigma}_s^-,\hat{\sigma}_{s'}^+]=-\hat{\sigma}_s^z\delta_{s,s'}$, we get
\begin{eqnarray}
\hat{\mathcal{H}}_k\ket{\ell} & = & \dfrac{\tilde{J}_k}{\sqrt{L}} \sum_{j,j'=1}^{L} e^{\frac{i2\pi\ell j'}{L}} \big( \! - \! \hat{\sigma}_j^+ \hat{\sigma}_{j'}^z \ket{0} \delta_{j',f(j+k)} \nonumber \\
&& \hspace{3.cm} - \hat{\sigma}_{f(j+k)}^+ \hat{\sigma}_{j'}^z \ket{0} \delta_{j,j'} \big) 
\label{eq:Ham2} \\
& = & 
\dfrac{\tilde{J}_k}{\sqrt{L}} \sum_{j=1}^{L} \Big( e^{\frac{i2\pi\ell f(j+k)}{L}}\hat{\sigma}_j^+\ket{0} + e^{\frac{i2\pi\ell j}{L} }\hat{\sigma}_{f(j+k)}^+ \ket{0} \Big).
\nonumber
\end{eqnarray}
For integer values of $\ell$, the exponential $\exp{i2\pi\ell f(j+k) / L}=\exp{i2\pi\ell (j+k) / L}$. Thus, the action of the Hamiltonian on the state is
\begin{equation}
    \hat{\mathcal{H}}_k\ket{\ell} = \dfrac{\tilde{J}_k}{\sqrt{L}} \sum_{j=1}^{L}\left( e^{\frac{i2\pi\ell (j+k)}{L} }\hat{\sigma}_j^+\ket{0} + e^{\frac{i2\pi\ell j}{L} }\hat{\sigma}_{f(j+k)}^+ \ket{0} \right).
    \label{Hkell}
\end{equation}
As a last stage, we rewrite the second summation in Eq.~\eqref{Hkell} as
\begin{equation}
    \sum_{j=1}^{L-k} e^{\frac{i2\pi\ell j}{L}}\hat{\sigma}_{f(j+k)}^+ \ket{0} + \sum_{j=L-k+1}^{L} e^{\frac{i2\pi\ell j}{L}}\hat{\sigma}_{f(j+k)}^+ \ket{0},
\end{equation}
where the first term $\sum_{j=1}^{L-k}$ runs on $j$ such that $j+k \leq L$, while the second one $\sum_{j=L-k+1}^{L}$ runs on $j+k>L$. In the first term we redefine $j'=j+k$ and in the second we redefine $j'=(j+k)-L$ (we basically defined $j' = f(j+k)$), thus obtaining
\begin{align}
    & \sum_{j'=k+1}^{L} e^{\frac{i2\pi\ell (j'-k)}{L}}\hat{\sigma}_{j'}^+\ket{0} + \sum_{j'=1}^{k} e^{\frac{i2\pi\ell (j'+L-k)}{L}}\hat{\sigma}_{j'}^+ \ket{0} \nonumber \\ &
   = \sum_{j'=1}^{L} e^{\frac{i2\pi\ell (j'-k)}{L}}\hat{\sigma}_{j'}^+ \ket{0}. 
     \label{sump}
\end{align}
Finally, substituting the second summation in Eq.~\eqref{Hkell} with the last expression in~\eqref{sump}, we obtain
\begin{align}
        \hat{\mathcal{H}}_k\ket{\ell} & =  \sum_{j=1}^{L}\dfrac{\tilde{J}_k}{\sqrt{L}}\left( e^{\frac{i2\pi\ell (j+k)}{L}}\hat{\sigma}_j^+\ket{0} + e^{\frac{i2\pi\ell (j-k)}{L}}\hat{\sigma}_{j}^+ \ket{0} \right) \nonumber \\ & =
    \tilde{J}_k\left(e^{\frac{i2\pi\ell k}{L}}+e^{-\frac{i2\pi\ell k}{L}}\right)\ket{\ell}=2 \tilde{J}_k \cos\left(\frac{2\pi\ell k}{L}\right)\ket{\ell}.
    \label{eq:fin_kpos}
\end{align}

On the other hand, for $k<0$, we can first rewrite the function $f(j+k)$ as
\begin{equation}
    f(j+k)=
    \begin{cases}
        j-|k| \qquad \;\: \operatorname{if} \; j+k > 0, \\
        j-|k|+L \, \: \: \operatorname{if} \; j+k \leq 0.
    \end{cases}
\end{equation}
Then, following the same procedure as for $k>0$ and using again the fact that $\exp{i2\pi\ell f(j+k)/L}=\exp{i2\pi\ell (j+k)/L}$, we get the same expression for $\hat{\mathcal{H}}_k\ket{\ell}$ as in Eq.~\eqref{Hkell}.
The second summation term in such expression can be cast as
\begin{equation}
    \sum_{j=|k|+1}^{L} e^{\frac{i2\pi\ell j}{L}}\hat{\sigma}_{j-|k|}^+ \ket{0} + \sum_{j=1}^{|k|} e^{\frac{i2\pi\ell j}{L}}\hat{\sigma}_{j-|k|+L}^+ \ket{0},
    \label{eq:split}
\end{equation}
where we split the summation in $j>|k|$ and $j\leq |k|$ terms. When $j>|k|$ we have $j-|k|=j+k>0$ and so $f(j+k)=j+k=j-|k|$; when $j\leq |k|$ we have $j-|k| = j+k\leq 0$ and so $f(j+k)=j+k+L=j-|k|+L$. Now, in the first and second summations of~\eqref{eq:split}, we introduce respectively $j ' = j-|k|$ and $j ' = j-|k|+L$ and obtain
\begin{align}
&\sum_{j'=1}^{L-|k|} e^{\frac{i2\pi\ell (j '+|k|)}{L}}\hat{\sigma}_{j '}^+ \ket{0} + \sum_{j '=1-|k|+L}^{L} e^{\frac{i2\pi\ell (j ' + |k| - L)}{L} }\hat{\sigma}_{j '}^+ \ket{0}  \nonumber \\
&
=\sum_{j'=1}^{L} e^{\frac{i2\pi\ell (j'+|k|)}{L}}\hat{\sigma}_{j '}^+ \ket{0} = \sum_{j'=1}^{L} e^{\frac{i2\pi\ell (j'-k)}{L}}\hat{\sigma}_{j '}^+ \ket{0} ,
\label{eq:last}
\end{align}
where we used $k = -|k|$. Finally, plugging the last expression of~\eqref{eq:last} into the second summation of Eq.~\eqref{Hkell}, we get
\begin{align}
\hat{\mathcal{H}}_k\ket{\ell} & =  \dfrac{\tilde{J}_k}{\sqrt{L}} \sum_{j=1}^{L} \left( e^{\frac{i2\pi\ell (j+k)}{L}}\hat{\sigma}_j^+\ket{0} + e^{\frac{i2\pi\ell (j-k)}{L}}\hat{\sigma}_{j}^+ \ket{0} \right)\nonumber  \\ & = \tilde{J}_k\left(e^{\frac{i2\pi\ell k}{L}}+e^{-\frac{i2\pi\ell k}{L}}\right)\ket{\ell}=2 \tilde{J}_k \cos\left(\frac{2\pi\ell k}{L}\right)\ket{\ell}.
\end{align}
Summarizing, a comparison between this expression and Eq.~\eqref{eq:fin_kpos} shows that the result of $\hat{\mathcal{H}}_k\ket{\ell}$ for negative values of $k$ is the same as that for positive $k$.

We now consider the full Hamiltonian
\begin{equation}
\hat{\mathcal{H}}_0 = \sum_{i\neq j}\tilde{J}_{ij} \big( \hat{\sigma}_i^+ \hat{\sigma}_j^- + {\rm H.c.} \big) = \sum_{i\neq j} \tilde{J}_{|i-j|} \big( \hat{\sigma}_i^+ \hat{\sigma}_j^- + {\rm H.c.} \big)
\end{equation}
Using the fact that $\tilde{J}_{ij}\sim \sin^{-3}(\pi |i-j| / L )=\sin^{-3}(\pi (L-|i-j|) / L )$, meaning that $\tilde{J}_{ij}=\tilde{J}_{|i-j|}=\tilde{J}_{L-|i-j|}$, it can be rewritten as 
\begin{equation}
\hat{\mathcal{H}}_0 = \sum_{k}\phantom{}^{'} \tilde{J}_k \sum_i \left( \hat{\sigma}_i^+\hat{\sigma}_{f(i+k)}^- + \rm H.c. \right)=\sum_{k}\phantom{}^{'} \hat{\mathcal{H}}_k.  
\end{equation}
Thus, the energy associated to the state $\ket{\ell}$, for a ring composed of $L$ atoms, is 
\begin{equation}
    E_{\ell}=2 \sum_{k}\phantom{}^{'}  \Tilde{J}_k \cos\left(\dfrac{2\pi\ell k}{L} \right) = 4\sum_{k}\phantom{}^{'}J_k \cos\left(\dfrac{2\pi\ell k}{L} \right).
\end{equation}

We conclude by observing that the current state $\ket{\ell}$ is also an eigenstate of the current operator. This can be proved using the same reasoning as before: we first apply the current operator in the ring on the current state
\begin{eqnarray}
    \hat{\mathcal{I}}\ket{\ell} & = & -i\dfrac{J_{n.n.}}{L\sqrt{L}}
    \sum_{j,j'=1}^L e^{\frac{i2\pi\ell j'}{L}} \big( \hat{\sigma}_j^+\hat{\sigma}_{f(j+1)}^- - {\rm H.c.} \big)\hat{\sigma}_{j'}^{+}\ket{0}
    \nonumber \\
    & = & 
    -i\dfrac{J_{n.n.}}{L\sqrt{L}} \bigg\{ \bigg[ \sum_{j=1}^{L-1}\left( e^{\frac{i2\pi\ell f(j+1)}{L}} \hat{\sigma}_j^+ - e^{-\frac{i2\pi\ell j}{L}} \hat{\sigma}_{f(j+1)}^+ \right) \! \bigg] \nonumber \\
    &&
    \qquad \; + e^{\frac{i2\pi\ell f(L+1)}{L}} \hat{\sigma}_L^+ - e^{-\frac{i2\pi\ell L}{L}} \hat{\sigma}_{f(L+1)}^+ \bigg\} \ket{0}
    \nonumber \\
    & = &
    -i\dfrac{J_{n.n.}}{L\sqrt{L}}\bigg\{ \bigg[ \sum_{j=1}^{L-1} \left( e^{\frac{i2\pi\ell j}{L}} e^{\frac{i2\pi\ell}{L}}  \hat{\sigma}_j^+ - e^{-\frac{i2\pi\ell j}{L}} \hat{\sigma}_{j+1}^+ \right) \! \bigg] \nonumber\\ 
    && \qquad \; + e^{\frac{i2\pi\ell L}{L}} e^{\frac{i2\pi\ell}{L}}\hat{\sigma}_L^+ - e^{\frac{i2\pi\ell}{L}} e^{-\frac{i2\pi\ell}{L}} \hat{\sigma}_{1}^+ \bigg\} \ket{0} \!,
\end{eqnarray}
where, in the last term of the summation, we simply used $e^{i 2 \pi \ell}=e^{i2\pi\ell/ L} e^{-i2\pi\ell / L}$, valid for integer $\ell$. Therefore, the final result is
\begin{equation}
    \hat{\mathcal{I}}\ket{\ell} =  -i\dfrac{J_{n.n.}}{L\sqrt{L}} \left( e^{\frac{i2\pi\ell}{L}} \! - \! e^{-\frac{i2\pi\ell}{L}} \right) \! \sum_{j=1}^{L} e^{\frac{i2\pi \ell j}{L}} \hat{\sigma}_j^+ \! \ket{0} ,
\end{equation}
that is,
\begin{equation}
     \hat{\mathcal{I}}\ket{\ell} = \dfrac{2J_{n.n.}}{L}\sin\left(\dfrac{2\pi\ell}{L} \right)\ket{\ell}.
\end{equation}
Therefore, the set of single-excitation current states $\{\ket{\ell}\}$ is a set of eigenstates for both current and Hamiltonian, thus the bare Hamiltonian and the current operator projected in the single excitation sector commute, the current is conserved when the system evolves under the bare Hamiltonian.

\section{Detuning patterns for different time-spacing}
\label{app:B}

\begin{figure}[!t]
\centering
\includegraphics[width=\columnwidth]{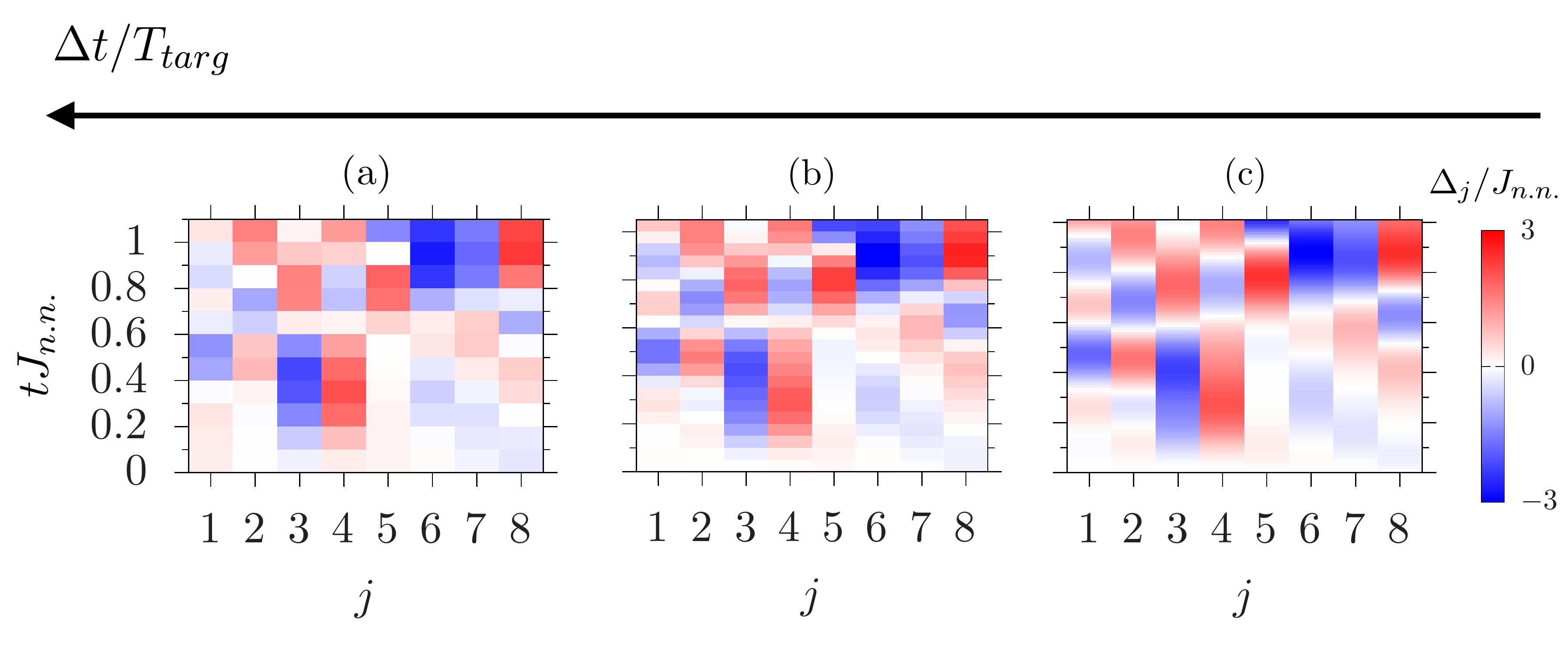}
\caption{Detuning profiles in time and position obtained through QOC protocol. The profiles are obtained fixing $T_{\rm targ}=J_{n.n.}^{-1}$ and $L=8$. The three panels correspond to different values of the time-spacing: $\Delta t / T_{\rm targ}=0.1$ (a), $\Delta t / T_{\rm targ}=0.05$ (b) and $\Delta t / T_{\rm targ}=0.01$ (c). The target state is $\ket{\ell=1}$.}
\label{fig:detnings_deltaT}
\end{figure}

In the main text we reported the detuning profiles extracted from QOC protocols for fixed $\Delta t$. Here, we report the full detuning profiles until the target time for different values of $\Delta t$ when the target state is the single current state $\ket{\ell=1}$ [Fig.~\ref{fig:detnings_deltaT}].

%%%%%%%%%%%%%%%%%%%%%%%%%%%%%%%%%
%%%%%%%%%%%%%%%%%%%%%%%%%%%%%%%%%

From Fig.~\ref{fig:detnings_deltaT} we observe that the detuning pattern needed to realize the $\ket{\ell=1}$ presents a complex behavior of alternating positive and negative detunings. In each site the detuning profiles are characterized by oscillations of reasonably small frequency. For this reason, increasing the length of time pulses, the fidelity remains close to one at the target time [see Fig.~\ref{fig:Fid_ell1_}(b)]. Indeed, the profile of the detuning for $\Delta t / T_{\rm targ}=0.1$ is basically a more discretized version of the case $\Delta t / T_{\rm targ}=0.01$, the similarity between the two is evident, indeed the fidelity is close to one at the target time for both cases.

\section{Quantum Speed Limit}
\label{app:QSL}
The Quantum Speed Limit (QSL) is the minimal time required to evolve between two orthogonal states under a time-independent Hamiltonian $\hat{\mathcal{H}}$. It can be computed as~\cite{deffner2017quantum, giovannetti2003quantumlimits} 
\begin{equation}
    \tau_{\rm QSL}=\max \left[\dfrac{\pi}{2\Delta E}, \dfrac{\pi}{2(E - E_0)} \right] ,
    \label{eq:QSL}
\end{equation}
where
\begin{equation}
E = \braket{\hat{\mathcal{H}}}, \qquad \Delta E = \sqrt{\braket{\hat{\mathcal{H}}^2} - \langle \hat{\mathcal{H}} \rangle^2},
\end{equation}
and $E_0$ is the ground-state energy of the system. Equation~\eqref{eq:QSL} denotes the maximum between the Mandelstam-Tamm time $\tau_{\rm MT}=\pi/(2\Delta E)$ and the Margolus–Levitin time $\tau_{\rm ML}=\pi/[2(E - E_0)]$~\cite{mandelstam1945uncertainty, margolus1998maximum, delcampo2013quantum}.

Given an initial state written in the basis of the Hamiltonian eigenstates
\begin{equation}
    \ket{\psi(0)}=\sum_n c_n \ket{E_n},
\end{equation}
the time-evolved state is
\begin{equation}
    \ket{\psi(t)}=\sum_n c_n e^{-iE_n t} \ket{E_n}.
\end{equation}
Its average energy and the corresponding variance 
\begin{equation}
E = \sum_n |c_n|^2 E_n,
\quad
(\Delta E)^2 = \tfrac{1}{2}\sum_{n,m}|c_n|^2|c_m|^2 (E_n - E_m)^2,
\end{equation}
remain unchanged during the time evolution. In the case of a superposition of two energy eigenstates $\ket{\ell}$ and $\ket{\ell '}$, we get
\begin{equation}
E = \tfrac{1}{2} (E_{\ell}+E_{\ell '}),
\quad
\Delta E = \tfrac{1}{2} (E_{\ell}-E_{\ell '}),
\end{equation}
with $E_{\ell}>E_{\ell '}$. We observe that
\begin{equation}
\Delta E \leq \tfrac{1}{2} ((E_{\ell}-E_0)+(E_{\ell '}-E_0)) = E - E_0,
\end{equation}
%\begin{align}
%\Delta E & = \tfrac{1}{2} ((E_{\ell}-E_0)-(E_{\ell '}-E_0)) \nonumber \\ & \leq \tfrac{1}{2} ((E_{\ell}-E_0)+(E_{\ell '}-E_0))=E - E_0,
%\end{align}
which implies $\tau_{\rm MT} \geq \tau_{\rm ML}$, thus $\tau_{\rm QSL}=\tau_{\rm MT}$ for the quantum superposition of two current states.

\section{Robustness of the protocol}
\label{app:rob}

The protocol we use in this work is based on Quantum Optimal Control optimization of the detuning profiles using the GRadient Ascent Pulse Engineering (GRAPE) algorithm~\cite{goerz2023grape, goerz2023quantumcontrol}. Given an initial state, a target state, and a time-dependent Hamiltonian, the algorithm is responsible for the engineering of a state that is as close as possible to the target state in a certain target time. The time-dependence of the Hamiltonian is discretized, the state is engineered through a gradient-descent minimization of the infidelity between target and the time-evolved states at the target time. In our case, the optimization parameters are local detunings $\Delta_j (t)$. 
Given a list of $N_T + 1$ times $t$ between $0$ and the target time $T_{\rm targ}$ separated by a certain $\Delta t$ time-interval, we obtain $N_T + 1$ different values of optimal detunings $\Delta_j(t)$. Once the optimal parameters are found, we evolve the system using a time-dependent Hamiltonian~\cite{Kramer2018quantum} in which at each time $t$ we start a pulse of length $\Delta t$. Considering that the physical pulses have a finite duration $\Delta t$, the detunings are switched off at $T_{\rm targ}+\Delta t$, which is approximately $T_{\rm targ}$ for sufficiently small $\Delta t$. 

The robustness of the ptotocol monitored through the Uhlmann fidelity~\cite{nielsen2012quantum} 
\begin{equation}
F(\rho,\sigma)=\Tr{\sqrt{\sqrt{\rho}\sigma\sqrt{\rho}}}^2,
\end{equation}
$\rho$ and $\sigma$ being two generic quantum states. In the case of pure states $\rho=\ket{\psi_{\rho}}\bra{\psi_{\rho}}$ and $\sigma=\ket{\psi_{\sigma}}\bra{\psi_{\sigma}}$ is reduces to
\begin{equation}
F(\rho,\sigma)=|\braket{\psi_{\rho}|\psi_{\sigma}}|^2.
\end{equation}
In this section we focus on the fidelity between a certain target state and the time evolved state. Through its dynamics, we can study the robustness of the protocol by changing the time parameters $\Delta t$, $T_{\rm targ}$, by the possible presence of noise in $\Delta_j(t)$ and by decoherence.

\subsection{Optimal target time}

%%%%%%%%%%%%%%%%%%%%%%%%%%%%%%%%%
\begin{figure}[!t]
\includegraphics[width=0.233\textwidth]{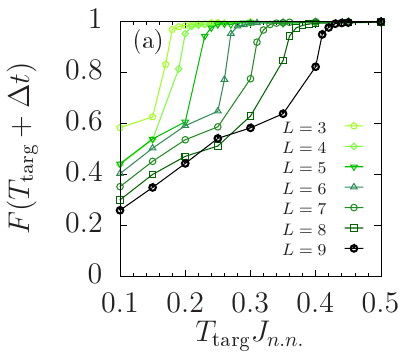}
\includegraphics[width=0.23\textwidth]{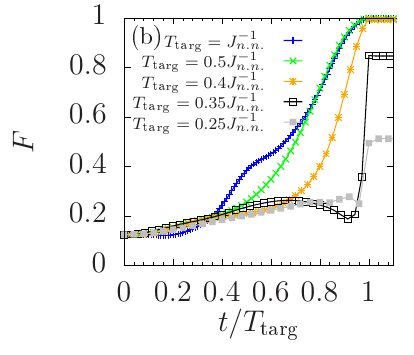}
\includegraphics[width=0.23\textwidth]{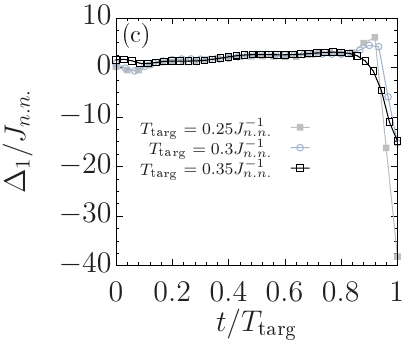}
\includegraphics[width=0.23\textwidth]{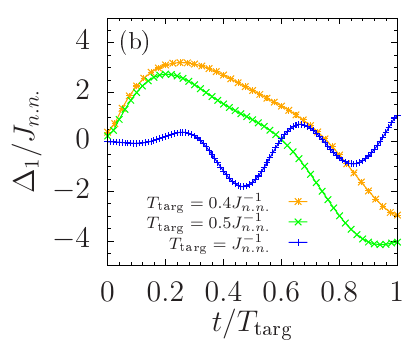}
\caption{(a) The fidelity between the target state $\ket{\ell\!=\!1}$ and the time evolved state at fixed time $T_{\rm targ} + \Delta t$, as a function of the target time, with $\Delta t = 0.01J_{n.n.}^{-1}$. Different ring sizes are considered. (b) The time evolution of the fidelity between the time evolved state and the target state $\ket{\ell=1}$ for different values of the target time, once $J_{n.n.}$ is fixed. We choose $\Delta t = 0.01J_{n.n.}^{-1}$ and $L=8$. (c-d) optimal time profiles of the detuning $\Delta_1$ on the first site, for different choices of the target time and fixed $\Delta t = 0.01J_{n.n.}^{-1}$, $L=8$. (c-d) The detuning profile for (c) small $T_{\rm targ}$, insufficent to reach a fidelity close to one at the target time, and for (d) $T_{\rm targ}$ sufficiently big to reach a fidelity close to one.}
\label{fig:Fid_target_time}
\end{figure}
%%%%%%%%%%%%%%%%%%%%%%%%%%%%%%%%%

%%%%%%%%%%%%%%%%%%%%%%%%%%%%%%%%%
\begin{figure*}[!t]
\centering
\includegraphics[width=\textwidth]{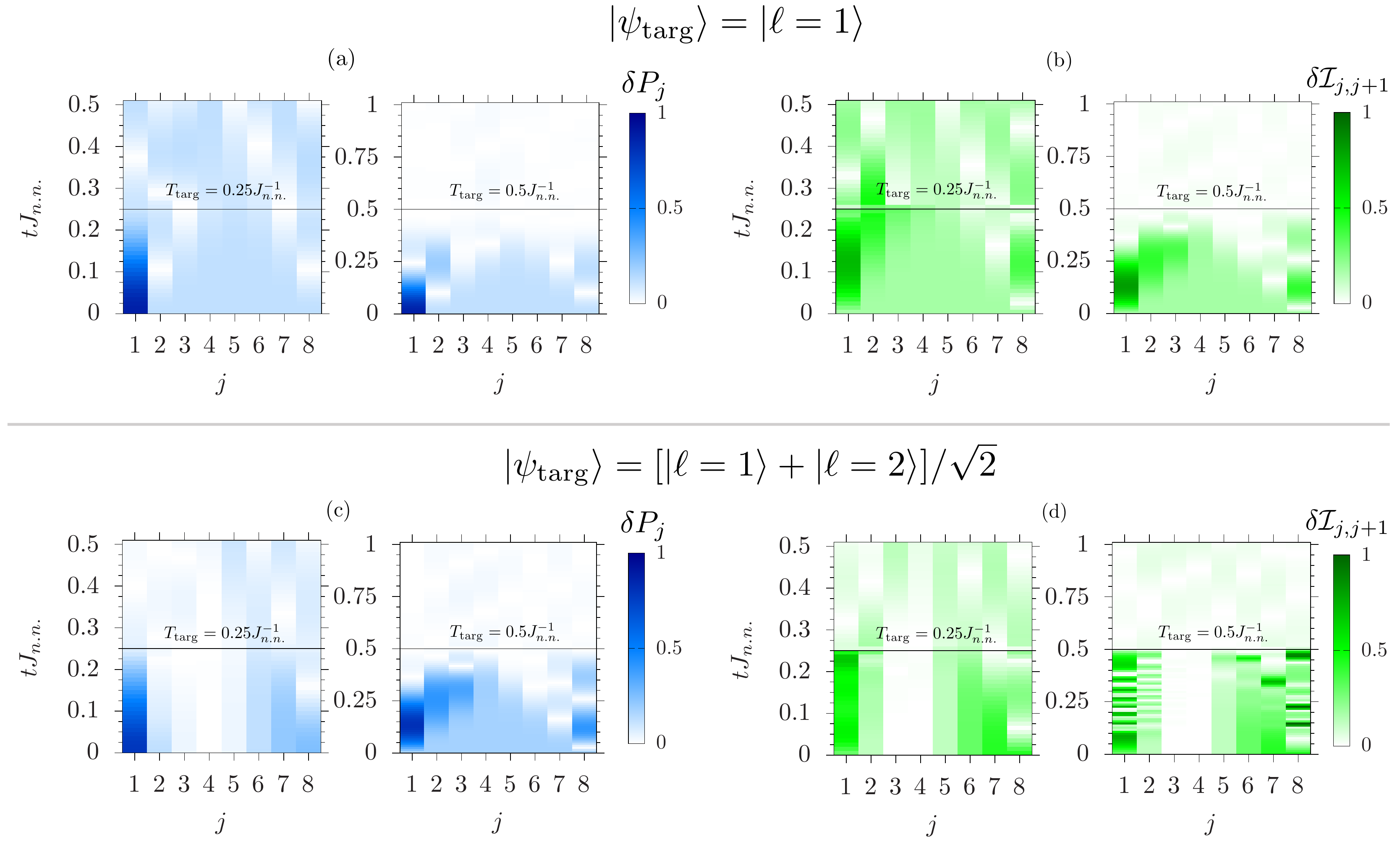}
\caption{(a-c) Local populations and (b-d) current errors, as defined in Eqs.~\eqref{eq:errors}, for different choices of target states and time. In particular we consider the target states (a-b) $\ket{\ell=1}$ and (c-d) $\tfrac{1}{\sqrt{2}}[\ket{\ell=1}+\ket{\ell=2}]$. We fix $\Delta t =0.01 J_{n.n.}^{-1}$ and $L=8$.}
\label{fig:errors_pop_curr}
\end{figure*}
%%%%%%%%%%%%%%%%%%%%%%%%%%%%%%%%%

To engineer a given target state, we need to understand what is the minimal target time $T_{\rm targ}^{(\rm min)}$ needed to reach the target state with good fidelity. Let us choose, as a target state, the single quantum current state $\ket{\ell\!=\!1}$:
\begin{equation}
\ket{\psi_{\rm targ}}=\dfrac{1}{\sqrt{L}}\sum_{j=1}^{L}e^{\frac{i2\pi j}{L}} \hat{\sigma}_j^+\ket{0} ,
\end{equation}
where $\ket{0}=\ket{\downarrow,\downarrow,\ldots,\downarrow}$. The initial state is the single localized excitation state $\ket{\uparrow,\downarrow,\ldots,\downarrow}$.  %We optimize the local control parameters to reach $\ket{\ell=1}$.
We compute the fidelity
\begin{equation}
F(t)=|\braket{\psi(t)|\psi_{\rm targ}}|^2,
\end{equation}
paying attention to the results at $T_{\rm targ} + \Delta t$.
Figure~\ref{fig:Fid_target_time}(a) displays the behavior of the fidelity between $\ket{\psi_{\rm targ}}$ and $\ket{\psi(T_{\rm targ}+\Delta t)}$ as a function of $T_{\rm targ}$ for different system sizes, which exhibits a jump. In fact, one can identify a time $T_{\rm targ}^{(\rm min)}$ that splits the plot in two qualitatively distinct regions. For $T_{\rm targ} < T_{\rm targ}^{(\rm min)}$, the fidelity is far from one and depends on $T_{\rm targ}$. For $T_{\rm targ} > T_{\rm targ}^{(\rm min)}$, the fidelity is not anymore dependent on the target time and it saturates to a value close to one. Since the target state is a coherent superposition of excitations with the same weight in each site of the ring, the minimal target time $T_{\rm targ}^{(\rm min)}$ depends on the system size: the initial localized excitation must explore the whole ring before reaching the target state, thus the required time increases with the size. The minimal target time growth in the size $L$ is approximately linear for $L \geq 5$.

In Fig.~\ref{fig:Fid_target_time}(b) we analyze the time behavior of the fidelity for different values of the target time, both for
$T_{\rm targ} > T_{\rm targ}^{(\rm min)}$ ($T_{\rm targ} = J_{n.n.}^{-1}, \, 0.5J_{n.n.}^{-1}, \, 0.4J_{n.n.}^{-1}$) and for $T_{\rm targ} < T_{\rm targ}^{(\rm min)}$ ($T_{\rm targ} = 0.35J_{n.n.}^{-1}, \,0.25J_{n.n.}^{-1}$).
In the former cases, the fidelity gently grows and reaches a value close to one at the target time. In the latter cases, the target time is too small to reach the target state with reasonable fidelity: The fidelity is small and remains far from one for most of the time, while, close to the target time, it suddenly increasing to a value that is still far from one. In Figs.~\ref{fig:Fid_target_time}(c-d) we report the time behavior of the detuning profile on the first site, for (c) $T_{\rm targ} < T_{\rm targ}^{(\rm min)}$ and (d) $T_{\rm targ} > T_{\rm targ}^{(\rm min)}$. For sufficiently big values of the target time, the detuning has a regular behavior in time, its values are comparable with those of the hopping strength. If the target time is small, the detuning becomes extremely irregular close to the target time, with visible sudden jumps, and the final state stays far from the desired one.

Given a ring with the number of atoms considered so far and the reference value of the nearest neighbor hopping strength $J_{n.n.}\sim 1$MHz, a target time of $0.5\mu$s is sufficient to reach a single current state with high fidelity using pulses of duration $\Delta t \sim 10$ns.

We finally study the effect of different choices of target time on the dynamics of local observables, focusing on the local population and the current. The distribution of the latter gives information on the phase pattern of the state. Indeed, given the generic one-excitation state $\ket{\psi}=\sum_{j=1}^L \sqrt{P_j} e^{i\phi_j}\hat{\sigma}^+_j\ket{0}$, the local excitation current reads
\begin{eqnarray}
\braket{\psi|\hat{\mathcal{I}}_{j,j+1}|\psi} & =& -iJ_{n.n.}\braket{\psi|\hat{\sigma}_{j}^+\hat{\sigma}^-_{j+1} - \rm H.c.|\psi} \\
& = & 2J_{n..n.} \big| \sqrt{P_j} \big| \, \big|\sqrt{P_{j+1}} \big| \sin\left(\phi_{j+1}-\phi_j\right) \nonumber
\end{eqnarray}
and so it is directly related to the local phases of state. We evaluate the local discrepancy between the state obtained through QOC and the target state using 
\begin{align}
    &\delta P_j = \left|\braket{\psi(t)|j}\braket{j|\psi(t)}-\braket{\psi_{\rm ex}(t)|j}\braket{j|\psi_{\rm ex}(t)}\right|,
    \\
    &\delta \mathcal{I}_{j,j+1} = \big| \! \braket{\psi(t)|\hat{\mathcal{I}}_{j,j+1}|\psi(t)}-\braket{\psi_{\rm ex}(t)|\hat{\mathcal{I}}_{j,j+1}|\psi_{\rm ex}(t)} \! \big|,   
    \label{eq:errors}
\end{align}
which we define, respectively, as the local population and current errors. The state $\ket{\psi(t)}$ is the one evolved under $\hat{\mathcal{H}}_0 + \sum_j \Delta_j(t)\hat{\sigma}_j^z$ until $\tilde{t}=T_{\rm targ}+\Delta t$, and under $\hat{\mathcal{H}}_0$ after $\tilde{t}$. Finally, we introduce the exact time-evolved state $\ket{\psi_{\rm ex}(t)}$ as the evolved of $\ket{\psi_{\rm targ}}$ under 
\begin{equation}
    \hat{\mathcal{H}}_{\rm ex}(t) = 
    \begin{cases}
        \hat{\mathbb{1}} \quad \; \; \textrm{if} \; \; t<\tilde{t}, \\
        \hat{\mathcal{H}}_0 \; \; \; \textrm{if} \; \; t\geq\tilde{t}.
    \end{cases}   
\end{equation}
$\hat{\mathcal{H}}_{\rm ex}(t)$ is responsible for the evolution of the ideal target state under $\hat{\mathcal{H}}_0$ after the time in which the optimization protocol ends, the latter is compared with those of the state obtained through the optimization protocol. 

Figure~\ref{fig:errors_pop_curr} reports the errors for different target times and target states; we consider a single current state $\ell=1$ and superposition of $\ell=1,2$. Panel (a) shows a comparison of the population error for two values of the target time and the same single current $\ell=1$ target state. The plot confirms that $T_{\rm targ}=0.25 J_{n.n.}^{-1}$ is not sufficient to realize with good approximation the target state, since $T_{\rm targ}<T_{\rm targ}^{(\rm min)}$. The error is close to zero only on two sites at the target time, elsewhere is considerably far from zero. In particular, the sites opposite to the initial excitation $j=4,5,6$ present population error far from zero at the target time; in this sites, the error is almost unchanged with respect to its initial value, it means that the excitation does not have sufficient time to reach the central sites. For $T_{\rm targ}=0.5 J_{n.n.}^{-1}$ ($T_{\rm targ}>T_{\rm targ}^{(\rm min)}$) the local error clearly decreases at the target time. The excitation can explore the whole ring during the dynamics and the population is almost homogeneous at the target time. Panel (b) reports the behavior of the local current error for the same single current target state. $T_{\rm targ}=0.25 J_{n.n.}^{-1}$ is not sufficient to generate the desired phase pattern in the ring through local detunings. As expected, for $T_{\rm targ}=0.5 J_{n.n.}^{-1}$ the local current error is close to zero at the target time. 

Panels (c) and (d) show a similar analysis, but for a target state being a superposition of $\ell=1$ and $\ell=2$ states. As in the single current state, the errors after the target time are considerably small for $T_{\rm targ}=0.5 J_{n.n.}^{-1}$. The main difference is observed for $T_{\rm targ}=0.25 J_{n.n.}^{-1}$: the target state presents a nonlocal distribution of population and current, the initial localized excitation does not need to explore the whole ring to reach the desired state. Therefore the errors on population and local currents for $T_{\rm targ}=0.25 J_{n.n.}^{-1}$ at the target time are smaller than in the case of single current state. Starting from a fully localized state, a quantum superposition of currents whose population distribution is peaked in a region close to the initial excitation can be realized faster than a single current state.

\subsection{Fidelity for a fixed target time}

%%%%%%%%%%%%%%%%%%%%%%%%%%%%%%%%%
\begin{figure}[!t]
\includegraphics[width=0.49\columnwidth]{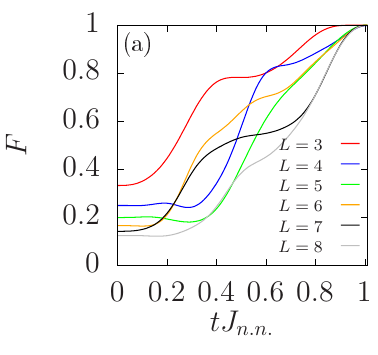}
\includegraphics[width=0.475\columnwidth]{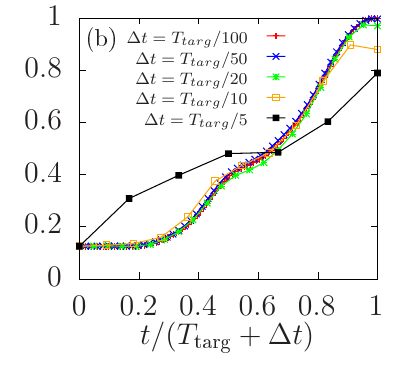}
\caption{Fidelity between the time-evolved state and the target state $\ket{\ell=1}$. Results are shown for (a) different values of $L$ and fixed time spacing $\Delta t = T_{\rm targ} / 100$, until $T_{\rm targ}=J_{n.n.}^{-1}$, and (b) different time spacings $\Delta t$ and $L=8$, with $T_{\rm targ}=J_{n.n.}^{-1}$ fixed.}
\label{fig:Fid_ell1_}
\end{figure}
%%%%%%%%%%%%%%%%%%%%%%%%%%%%%%%%%

%%%%%%%%%%%%%%%%%%%%%%%%%%%%%%%%%
\begin{figure*}[!t]
\hfill
\subfigure[]{\includegraphics[width=5.85cm]{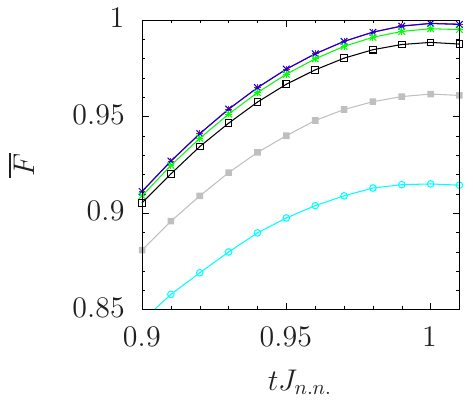}}
\hfill
\subfigure[]{\includegraphics[width=6.05cm]{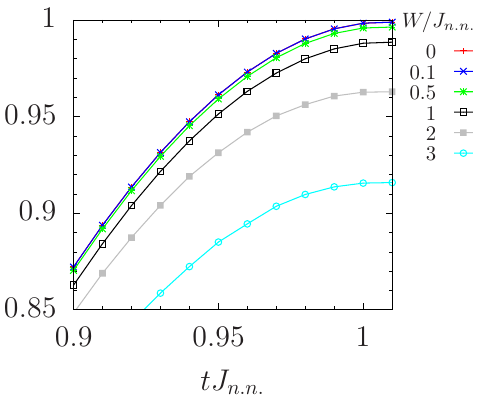}}
\hfill
\subfigure[]{\includegraphics[width=5.32cm]{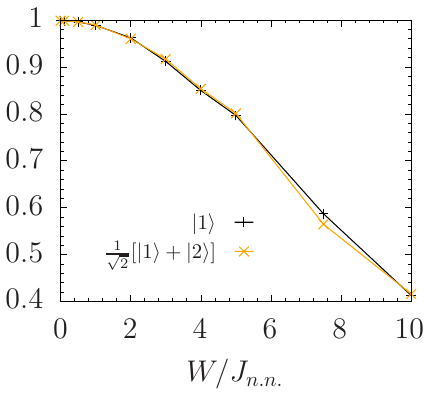}}
\hfill
\caption{The fidelity between the target and the time-evolved state in the presence of noise in the detunings, averaged over $N_{\rm rea} = 100$ disorder realizations. Two different target states are considered: (a) $\ket{\ell=1}$ and (b) $\tfrac{1}{\sqrt{2}}[\ket{\ell=1}+\ket{\ell=2}]$. The dynamics is shown in a small time window around the target time. Panel (c) reports the same quantity at the target time, for both choices of target states (black and orange curves), as a function of $W$. We set $T_{\rm targ}=J_{n.n.}^{-1}$, $\Delta t = T_{\rm targ}/ 100$, and $L=8$.}
\label{fig:Fid_disorder}
\end{figure*}
%%%%%%%%%%%%%%%%%%%%%%%%%%%%%%%%%

We now consider different system sizes and pulse durations $\Delta t$, keeping $T_{\rm targ}=J_{n.n.}^{-1}$ fixed. From the previous section, we know that this choice permits to reach the single current target state with good fidelity for $\Delta t = 0.01 J_{n.n.}^{-1}$. We monitor the fidelity dynamics between time-evolved and target state.

Figure~\ref{fig:Fid_ell1_}(a) reports the behavior of the fidelity for different sizes, fixing a pulse duration $\Delta t = T_{\rm targ}/100$ (if $T_{\rm targ}=1\mu$s, we have $\Delta t=10$ns). The fidelity starts from $1/L$ (that is, the smallest value) and reaches a value close to one, for any choice of $L$.  In all cases, $T_{\rm targ}>T_{\rm targ}^{(\rm min)}$; this is the reason why the fidelity is very close to one at the target time. Looking at the time evolution, we are not able to recognize a regular dependence on the system size: while $L=3$ presents the largest fidelity at almost any time, for other sizes we observe many crossings between the different curves, indicating an irregular behavior.

Conversely, Fig.~\ref{fig:Fid_ell1_}(b) is for fixed $L=8$ and for different values of the time-intervals $\Delta t$.  The optimization protocol ends at time $T_{\rm targ} + \Delta t$, thus we normalize the times as $t/(T_{\rm targ} + \Delta t)=t/(T_{\rm targ} + T_{\rm targ}/N_T)=tJ_{n.n.}/(1 + 1/N_T)$. For $\Delta t \in [0.01,0.05]T_{\rm targ}$, the behavior of the fidelity is quite similar and it reaches a value close to one at the end of the optimization protocol. The case $\Delta t = T_{\rm targ}/5$ is different from the other ones since the fidelity value at $T_{\rm targ}+\Delta t$ is far from one, being $F \lesssim 0.8$. We checked that $\Delta t = T_{\rm targ}/10$ works slightly better, despite at the end of the optimization protocol we are still left with significant discrepancies of the fidelity from one. 
Considering a reference value of $J_{n.n.}\sim 1$MHz for the nearest neighbor hopping, as well a typical target time $T_{\rm targ}\sim 1\mu$s, we have that pulses of duration $\Delta t \sim 50$ns work sufficiently well, while those of duration $\Delta t \sim 200$ns are not able to reach the desired state.

\subsection{Fidelity in the presence of noise}

Understanding how the dynamics is affected by possible noise in the detuning can be relevant under an experimental point of view. Here we study the dynamics of the fidelity between the target and the time-evolved state in the presence of noise in the detuning. To do so, we extract the detuning profiles $\Delta_j(t)$ from the QOC protocol, then we modify them as 
\begin{equation}
    \Delta_j(t) \rightarrow \Delta_j(t) + \varepsilon_j(t),
\end{equation}
where $\varepsilon_j(t)$ is a random number uniformly distributed in $[-W,W]$ and $W$ is the disorder strength. We note that $\varepsilon_j(t)$ is a $L\times (N_T+1)$ matrix, $N_T=T_{\rm targ}/\Delta t$ being total number of time intervals. 
We repeat this procedure for a number $N_{\rm rea}$ of disorder realization. In each of them we extract $F_{\beta}(t)=|\braket{\psi_{\beta}(t)|\psi_{\rm targ}}|^2$ [here $\beta=1,\ldots,N_{\rm rea}$ labels the realization and $\ket{\psi_{\beta}(t)}$ is the corresponding time-evolved state]. Then, we average over the disorder
\begin{equation}
\overline{F(t)}=\dfrac{\sum_{\beta=1}^{N_{\rm rea}}F_{\beta}(t)}{N_{\rm rea}}.
\end{equation}

Figure~\ref{fig:Fid_disorder} reports the behavior of such quantity, $\overline{F(t)}$, for different values of the disorder strength and for the two target states $\ket{\ell=1}$ and $\tfrac{1}{\sqrt{2}}[\ket{\ell=1}+\ket{\ell '=2}]$. In both cases the protocol is robust in the presence of noise in the detunings. Moreover, the disorder affects the protocol to reach a single current state and a superposition of current states in the same way, as is clearly visible from Fig.~\ref{fig:Fid_disorder}(c): It starts with a slow decrease for small values of $W$ and then it goes linearly for larger values. Note that, for large $W$, the two fidelities are not perfectly superposed, since the disorder acquires a more important role. To conclude, we find $\overline{F}>0.95$ at the target time for $0<W/J_{n.n.}\lesssim 2$, meaning that a disorder of magnitude comparable with the nearest-neighbor hopping is not particularly dangerous.

\subsection{Robustness under pure dephasing}
\label{app:puredeph}

Due to the sensitivity of Rydberg states to noise in the laser fields and fluctuations of the atom positions, decoherence can have effects on the system dynamics~\cite{morsch2018dissipative}. In the following, we probe the robustness of our results under pure dephasing. The crucial questions are: how effective is the QOC protocol under dephasing and how much the current may survive after the target time. Other sources of decoherence, like relaxation of the Rydberg states to the ground state, are expected to occur on times of the order of hundreds $\mu$s, much longer than those of the excitation transport. To this purpose, we study the nonunitary time evolution for the system's density
matrix $\rho$, ruled by a Markovian Lindblad master equation~\cite{breuer2007theory}: 
\begin{equation}
  \dfrac{\partial \rho}{\partial t}= -i[\hat{\mathcal{H}}(t),\rho] + \gamma \sum_{j=1}^L \mathcal{D}_\rho [\hat{\sigma}_j^{z}] ,
  \label{eq:Master_eq}
\end{equation}
with $\mathcal{D}_\rho[\hat{O}] = \hat{O}\rho \hat{O}^\dagger - \tfrac12 \{ \hat{O}^\dagger \hat{O}, \rho \}$ and $\gamma$ being the dephasing rate.

The expectation value of the nearest-neighbor current operator~\eqref{eq:current_op} is computed as $\mathcal{I} = \Tr \big[ \rho \, \hat{\mathcal{I}} \big] $, so that
\begin{align}
\dfrac{\partial \mathcal{I}}{\partial t } \! & = -i \Tr [[\hat{\mathcal{H}},\rho] \, \hat{\mathcal{I}}] + \gamma \sum_j \Tr [\big(\hat{\sigma}_j^z\rho\hat{\sigma}_j^z \!-\! \tfrac{1}{2}\{\hat{\sigma}_j^z\hat{\sigma}_j^z,\rho \} \big) \hat{\mathcal{I}}] \nonumber \\
& = -i \Tr [[\hat{\mathcal{I}},\hat{\mathcal{H}}] \, \rho] + \gamma \sum_j \Tr [ \big( \hat{\sigma}_j^z\hat{\mathcal{I}}\hat{\sigma}_j^z - \hat{\mathcal{I}} \big) \, \rho].
\end{align}
Switching off the detunings, one has
$[\hat{\mathcal{I}},\hat{\mathcal{H}}]=0$ and thus the current is conserved: as a consequence, its dynamics is governed only by the incoherent part
\begin{align}
\dfrac{\partial \mathcal{I}}{\partial t } & = \gamma \sum_j \left\{ \Tr \big[ \hat{\sigma}_j^z \, \hat{\mathcal{I}} \, \hat{\sigma}_j^z \, \rho \big] - \Tr \big[ \hat{\mathcal{I}} \, \rho \big] \right\} \nonumber \\
& = \gamma \sum_j \left\{ \Tr \big[ ( \hat{\sigma}_j^z)^2 \, \hat{\mathcal{I}}\rho \big] + \Tr \big[ \hat{\sigma}_j^z \, [\hat{\mathcal{I}},\hat{\sigma}_j^z] \, \rho \big] -\Tr \big[ \hat{\mathcal{I}} \, \rho \big] \right\} \nonumber \\
& = \gamma \sum_j \Tr \big[ \hat{\sigma}_j^z \,[ \hat{\mathcal{I}},\hat{\sigma}_j^z ] \, \rho \big].
\label{eq:current_deph}
\end{align}
We now compute the commutator
\begin{equation}
    \big[ \hat{\mathcal{I}},\hat{\sigma}_j^z \big] = \dfrac{-2iJ_{n.n.}}{L} \big( \hat{\sigma}^+_{j-1}\hat{\sigma}^-_{j} + \hat{\sigma}^-_{j-1}\hat{\sigma}^+_{j} -  {\rm H.c.} \big)
\label{eq:commutator_deph}
\end{equation}
and insert~\eqref{eq:commutator_deph} into the final expression in~\eqref{eq:current_deph} to obtain
\begin{equation}
\dfrac{\partial \mathcal{I}}{\partial t}= -4\gamma \Tr \big[ \rho \, \hat{\mathcal{I}} \big],
\label{eq:curr_diss_eq}
\end{equation}
where we used $\sum_j \hat{\sigma}^+_{j-1}\hat{\sigma}^-_{j} = \sum_j \hat{\sigma}^+_{j}\hat{\sigma}^-_{j+1}$ (valid in a ring) and $\hat{\sigma}^z_{j}\hat{\sigma}^-_{j}=-\hat{\sigma}^-_{j}$, $\hat{\sigma}^z_{j}\hat{\sigma}^+_{j}=\hat{\sigma}^+_{j}$.
Equation~\eqref{eq:curr_diss_eq} implies 
\begin{equation}
\mathcal{I} \sim e^{-4\gamma t}.
\label{eq:currdecay}
\end{equation}

%%%%%%%%%%%%%%%%%%%%%%%%%%%%%%%%%
\begin{figure}[!t]
\includegraphics[width=0.47\columnwidth]{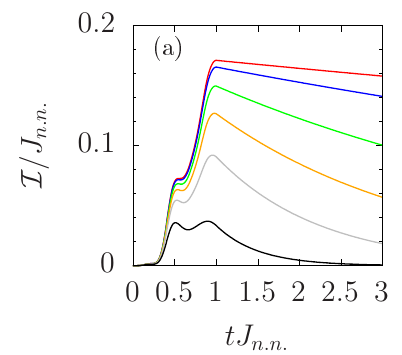}
\includegraphics[width=0.505\columnwidth]{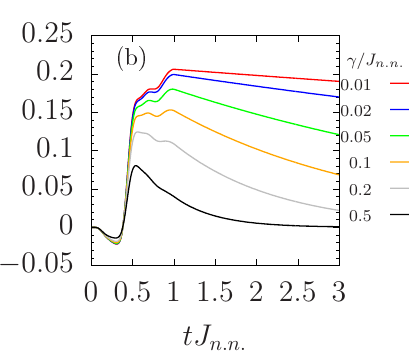}
\includegraphics[width=0.47\columnwidth]{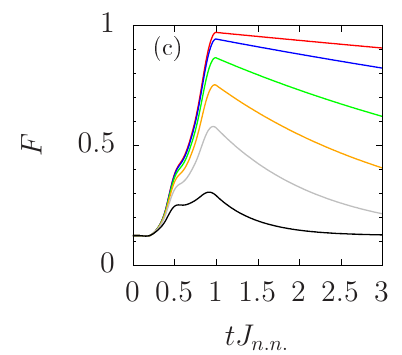}
\includegraphics[width=0.505\columnwidth]{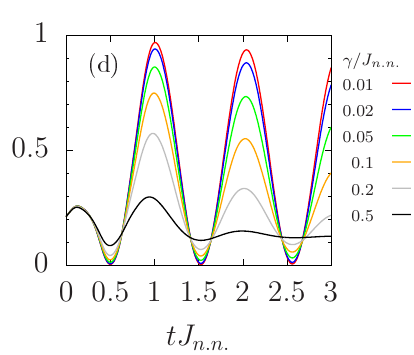}
\caption{The current and the fidelity dynamics under pure dephasing for the target states $\ket{\ell=1}$ [(a) and (c), respectively] and $\tfrac{1}{\sqrt{2}} [\ket{\ell=1}+\ket{\ell=2}]$ [(b) and (d), respectively], for different values of the dephasing rate. We fix $L=8$, $T_{\rm targ}=J_{n.n.}^{-1}$, and $\Delta t = T_{\rm targ}/100$.}
\label{fig:Idephasing}
\end{figure}
%%%%%%%%%%%%%%%%%%%%%%%%%%%%%%%%%

Figure~\ref{fig:Idephasing} shows the time behavior of the current, until and after the target time $T_{\rm targ}=J_{n.n.}^{-1}$. We extract the optimal detuning shape in the absence of dissipation, then we evolve the system under dissipation in order to study how the dephasing damages the flow. In panels (a) and (b) we focus on the current for two different target states: the one current state and the superposition of two current states, respectively. As expected from Eq.~\eqref{eq:currdecay}, after the target time, the current undergoes a decaying behavior with rate $4\gamma$. The presence of a nonzero current is related to the coherences of the state. Since pure dephasing destroys coherences, it also kills the current, being responsible for its exponential decay.

We also computed the fidelity between the time-evolved and the target state
\begin{equation}
F=\Tr{\sqrt{\sqrt{\rho(t)}\rho_{\rm targ}\sqrt{\rho(t)}}}^2
\end{equation}
where $\rho_{\rm targ}=\ket{\psi_{\rm targ}}\bra{\psi_{\rm targ}}$ and $\rho(t)$ is computed from \eqref{eq:Master_eq}.
Panels (c) and (d) report $F(t)$ for $M=1$ and $M=2$, respectively. In the case of a single current state, the fidelity reaches a values closer and closer to one at the target time, as $\gamma$ decreases, and then decays following a similar behavior of the current. In the case of a superposition of two current states, the fidelity behavior is different from that of the current. While the current decays after the target time, $F(t)$ shows characteristic oscillations due to the nontrivial excitation flow in the ring. As $\gamma$ increases, the amplitude of oscillations reduces, due to the suppression of the coherent excitation flow. In both cases, a dephasing rate $\gamma=0.5J_{n.n.}$ is sufficient to reach the steady state in the time-interval considered. The current and fidelity values in the steady state are $0$ and $1/L$, since this coincides with the maximally mixed state $\rho_{SS}=\tfrac{1}{L} \sum_{j}\ket{j}\bra{j}$, where $\ket{j}\equiv\hat{\sigma}_j^+\ket{0}$.

\section{Read-out of the state}

The typical state detection procedure in this type of systems is based on the coupling of one of the two Rydberg states (i.e., $\ket{\uparrow}$) with a low-lying state. The atoms in this latter state can be imaged, while the others are lost. Imaged atoms correspond to those originally in $\ket{\uparrow}$, while lost atoms correspond to thos originally in $\ket{\downarrow}$.
 In this way it is possible to reconstruct site-by-site the population of the system~\cite{chen2023spectroscopy, barredo2015coherent}. Using this procedure together with global rotations of the system, it is possible to measure populations in the $x,y$ basis and correlation functions $\braket{\hat{\sigma}_i^{\alpha} \hat{\sigma}_j^{\alpha}}$ ($\alpha=x,y,z$). Recently, the multi-basis correlation function measurement has been performed experimentally~\cite{bornet2024enhancing}, exploiting the local shifts induced by the addressing beams: local rotations using MW pulses resonant only with the sites of interest are performed, correlation functions of the form $\braket{\hat{\sigma}_i^{\alpha} \hat{\sigma}_j^{\beta} \ldots \hat{\sigma}_k^{\gamma}}$ can be in principle computed. Multi-basis measurements permit to experimentally access quantum currents
\begin{align}
 \braket{\psi(t)|\hat{\mathcal{I}}|\psi(t)} & = -i \frac{J_{n.n.}}{L} \sum_{j=1}^{L} \braket{\psi(t)|\hat{\sigma}_{j}^+\hat{\sigma}^-_{j+1} - \rm H.c.|\psi(t)} \nonumber \\
& = \frac{J_{n.n}}{2L}\sum_{j=1}^{L}\braket{\psi(t)|\hat{\sigma}_{j}^y\hat{\sigma}^x_{j+1} - \hat{\sigma}_{j}^x\hat{\sigma}^y_{j+1}|\psi(t)}.
\end{align}
Local currents $\mathcal{I}_j\propto \braket{\psi(t)|\hat{\sigma}_{j}^y\hat{\sigma}^x_{j+1} - \hat{\sigma}_{j}^x\hat{\sigma}^y_{j+1}|\psi(t)}$ are measurable as well.

In the case of a single current state $\ket{\ell}$, both the expected local populations $P_j=1/L$ on each site and the total current $\mathcal{I}=2 (J_{n.n.}/L) \sin \left(2\pi\ell/L \right)$ are accessible quantities. Moreover, given a generic state $\ket{\psi}=\sum_{j=1}^L \sqrt{P_j} e^{i\phi_j}\hat{\sigma}^+_j\ket{0}$, it is possible to prove that
\begin{equation} \braket{\psi|\hat{\sigma}_i^{x} \hat{\sigma}_j^{x}|\psi}=2\sqrt{P_i} \sqrt{P_j}\cos(\phi_i - \phi_j).
\end{equation}
Thus, through the measurable quantity $\phi_i - \phi_j=\cos^{-1}\left(    \braket{\psi|\hat{\sigma}_i^{x} \hat{\sigma}_j^{x}|\psi}/(2\sqrt{P_i} \sqrt{P_j})\right)$ one can have direct information on the phases associated to each site. Indeed, if we consider $j=i+1$, we have $L$ equations for $\phi_i - \phi_{i+1}$ in $L$ unknowns $\phi_i$.

As regards the quantum currents superposition state $\ket{\Psi_{\Lambda}}$
in Eq.~\eqref{eq:currentSuperpos}, we can write it in the form 
\begin{equation}
\ket{\Psi_{\Lambda}} = \sum_{j=1}^L \! \bigg( \mathcal{N}\sum_{\ell \in \Lambda} e^{i2\pi\ell j / L} \bigg) \hat{\sigma}^+_j\ket{0} = \sum_{j=1}^L \sqrt{P_j} e^{i\phi_j}\hat{\sigma}^+_j\ket{0}.
\end{equation}
Its population and phase pattern are experimentally accessible and can be compared with the analytical forms:
\begin{eqnarray}
    P_j & = & \bigg| \mathcal{N}\sum_{\ell \in \Lambda} e^{i2\pi\ell j / L} \bigg|^2 ,
    \\
    \phi_j & = & \tan^{-1} \left\{ \frac{\Im \big[ \sum_{\ell \in \Lambda} e^{i2\pi\ell j / L} \big] }{ \Re \big[ \sum_{\ell \in \Lambda} e^{i2\pi\ell j / L} \big] }\right\}.
\end{eqnarray}
Concerning the dynamics of such state, we exploit the fact that $\ket{\ell}$ is an eigenstate of the bare Hamiltonian, so that, after the target time, the state evolves as
\begin{equation}
\ket{\Psi_{\Lambda}(t)}=\sum_{j=1}^L \bigg( \mathcal{N}\sum_{\ell \in \Lambda} e^{i(2\pi\ell j / L-E_{\ell}t)} \bigg) \hat{\sigma}^+_j\ket{0} .  
\end{equation}
This permits to have access to local time dependent populations $P_j(t)$ and phases $\phi_j(t)$.
Other less specific quantities as the number of excitation blobs (or nodes) that move in the ring or the excitation velocity can help in having immediate information on the nature of the state.

\twocolumngrid

\end{document}